\documentclass{article}

\usepackage{arxiv}

\usepackage[utf8]{inputenc} 
\usepackage[T1]{fontenc}    
\usepackage{hyperref}       
\usepackage{url}            
\usepackage{booktabs}       
\usepackage{amsfonts}       
\usepackage{nicefrac}       
\usepackage{microtype}      
\usepackage{lipsum}		
\usepackage{graphicx}
\usepackage{natbib}
\usepackage{doi}
\usepackage{amsmath}

\title{Changes in Crime Rates During the COVID-19 Pandemic}

\author{ \href{https://orcid.org/0000-0003-3718-3405  }{\includegraphics[scale=0.06]{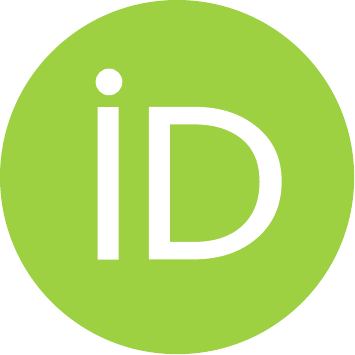}\hspace{1mm}Mikaela Meyer} \\
	Department of Statistics \& Data Science and Heinz College\\
	Carnegie Mellon University\\
	Pittsburgh, PA 15213 \\
	\texttt{mikaela@stat.cmu.edu} \\
	\And
	Ahmed Hassafy\\
	Heinz College\\
	Carnegie Mellon University\\
	Pittsburgh, PA 15213
    \And
	Gina Lewis \\
	Heinz College\\
	Carnegie Mellon University\\
	Pittsburgh, PA 15213
    \And
    Prasun Shrestha\\
    Heinz College\\
    Carnegie Mellon University\\
    Pittsburgh, PA 15213
    \And
    Amelia M. Haviland\\
    Heinz College\\
    Carnegie Mellon University\\
    Pittsburgh, PA 15213
    \And
    \href{https://orcid.org/0000-0003-0075-1851}{\includegraphics[scale=0.06]{orcid.pdf}\hspace{1mm}Daniel S. Nagin}\\
    Heinz College\\
    Carnegie Mellon University\\
    Pittsburgh, PA 15213
}

\renewcommand{\undertitle}{}

\hypersetup{
pdftitle={Changes in Crime Rates During the COVID-19 Pandemic},
pdfauthor={Mikaela Meyer et al.},
pdfkeywords={First keyword, Second keyword, More},
}

\begin{document}
\maketitle

\begin{abstract}
	\textbf{\textit{Research Summary}}: We estimate changes in the rates of five FBI Part 1 crimes—homicide, auto theft, burglary, robbery, and larceny—during the COVID-19 pandemic from March through December 2020. Using publicly available weekly crime count data from 29 of the 70 largest
cities in the U.S. from January 2018 through December 2020, three different linear regression model specifications are used to detect changes. One detects whether crime trends in four 2020 pre- and post-pandemic periods differ from those in 2018 and 2019. A second looks in more detail at the spring 2020 lockdowns to detect whether crime trends changed over successive biweekly periods into the lockdown. The third uses a city-level openness index that we created for the purpose of examining whether the degree of openness was associated with changing crime rates. For homicide and auto theft, we find significant increases during all or most of the pandemic. By contrast, we find significant declines in robbery and larceny during all or part of the pandemic and no significant changes in burglary over the course of the pandemic. Only larceny rates fluctuated with the degree of each city's lockdown.

\textbf{\textit{Policy Implications}}: It is unusual for crime rates to move in different directions, and the reasons for the mixed findings for these five Part 1 Index crimes—one with no change, two with sustained increases, and two with sustained decreases—are not yet known. We hypothesize that the reasons may be related to changes in opportunity, and the pandemic provides unique opportunities for future research to better understand the forces impacting crime rates. In the absence of a clear understanding of the mechanisms by which the pandemic affected crime, in the spirit of evidence-based crime policy, we caution against advancing policy at this time based on lessons learned from the pandemic ``natural experiment.''
\end{abstract}


\section{Introduction}
The COVID-19 pandemic has prompted a few early analyses on the impacts of lockdowns aimed at curtailing the pandemic’s spread on crime rates. The initial lockdowns began in March 2020 and generally eased in May 2020. Another momentous event, the death of George Floyd at the hands of Minneapolis police on May 25, 2020, coincided with the easing of the lockdowns. Floyd’s killing prompted protests across the US which were overwhelmingly peaceful but, in some cities, involved violence or property damage by protesters or counter-protesters.
During the summer and fall of 2020, various media sources reported that homicide counts had risen in some of America’s largest cities based on year-to-date comparisons (\citep{asher_its_2020}; \citep{hilsenrath_homicide_2020}; \citep{mccarthy_major_2020}; \citep{struett_chicago_2020}) and month-to-month comparisons \citep{campbell_violent_2020} between 2020 and previous years. Meanwhile, other crimes, such as burglary and robbery, were reported to have decreased in the months following the start of the COVID-19 pandemic in the U.S. (\citep{becker_violent_2020}, \citep{fbi_overview_2020}, \citep{lopez_rise_2020}).

In this paper, we report findings of an analysis of the pandemic’s impact in 29 of the 70 largest U.S. cities on five FBI Part 1 Index crimes—homicide, robbery, burglary, larceny, and auto theft. These are all of the cities that make weekly crime counts available. Concerning analyses of crime rates following Floyd’s killing, we are not able to disentangle the underlying reasons for changes we identify, which may include social unrest, economic conditions, changed police behavior, or other factors. Our aim instead is to carefully document changes or the absence of changes accounting for seasonal trends.  In March through May of 2020, the months following the onset of the pandemic but prior to George Floyd’s killing, we find burglary and auto theft rates did not significantly differ in 2020 compared to the same time of year in 2018 and 2019. Homicide rates increased, and robbery and larceny rates decreased during this pandemic lockdown time period in 2020 relative to the previous two years. We also analyzed how the amount of time that passed since each city went into lockdown was associated with changes in crime rates. This analysis found that in some biweekly periods during the lockdown, there were statistically significant changes in rates of auto theft (increases), robbery (reductions), and larceny (reductions) but not the other crime types. However, we also found that the city’s degree of lockdown strictness in each period was generally not associated with statistically significant crime rate changes. Specifically, we created an “openness index” that assigned integer values to cities each week based on the degree of the cities’ lockdowns; we found that only weekly larceny rates changed in concert with the openness index, the other crime types did not. 

In the summer months, which are traditionally associated with higher crime rates, we tested whether the seasonal changes in crime rates in 2020 were significantly different from trends in the previous two years or from any 2020 pre-pandemic change in crime rates, and whether any summer change persisted through the rest of 2020. We found that the average homicide rate in summer 2020 was significantly higher than the summer rates in 2018 and June 2019, and that this increase persisted through the rest of 2020. For auto theft, the analysis identified an increase in summer 2020, relative to seasonal trends in prior years, which again persisted throughout the rest of 2020. We also detected crime declines. As with during the lockdown period, robbery and larceny rates were lower than usual in summer 2020 and these declines persisted through the rest of 2020.  There was weak evidence of systematic declines in burglary in the summer and remainder of 2020 with these declines reaching only marginal statistical significance levels (less than 0.10). Thus, overall, we find that the COVID-19 pandemic and mass protests related to the killing of George Floyd have been associated with significant but mixed changes in urban crime in the U.S.

Due to the recency of these events, few academic articles have been written about how crime rates have changed during pandemic-related lockdowns and following the loosening of these lockdowns in the U.S.. \cite{piquero_staying_2020} analyzed how the COVID-19 lockdown in Dallas was associated with a spike in domestic violence in the two weeks after the lockdown began. \citep{campedelli_exploring_2020}also found mixed associations with the lockdown in Los Angeles and various crime rates in March 2020. \citep{mohler_impact_2020} analyzed daily police calls-for-service data from Los Angeles and Indianapolis from a month after lockdowns went in place and used regression models to compare these months to the months of 2020 leading up to the lockdowns. They found mixed results across crimes and the two cities. \citep{ashby_initial_2020} used seasonal auto-regressive integrated moving average (SARIMA) models to address seasonality while determining whether certain classes of crimes are significantly different from these models’ forecasts in sixteen large U.S. cities. Each city was modeled separately, and the data coverage ended on May 10th. This early analysis found no evidence in this set of cities that serious assaults in public or in residences increased or decreased between when the first COVID-19 case was reported in the U.S. in January until May 10th. The results also suggested that changes in crime trends varied across these cities. \citep{abrams_covid_2020} also provided an early analysis of a variety of crime rates per capita. Using data from 23 U.S. cities, this paper found that there was no statistically significant increase in homicides the four weeks after lockdowns went in place. However, \citep{abrams_covid_2020} found that drug crimes, violent crimes, and most property crimes declined in these four weeks in those cities. Finally, in their report for the Council on Criminal Justice, \cite{rosenfeld_pandemic_2020} 
look for structural breaks in the rate of change of weekly crime rates aggregated over at most 24 U.S. cities, depending on the crime type, between January 2017 and November 2020. Using this method, they find that homicide and motor vehicle theft were higher in spring and summer 2020 than in 2019, though larceny and residential burglaries declined in this period, and robbery did not see a significant rate change. 

The difference between the findings of our study and some previous analyses of homicide and other crime rates is likely attributable to several factors. First, we analyze data through December 2020 from 29 cities. Unlike other analyses, our focus was not on identifying individual cities that might have experienced increases in crime; rather it was to examine whether in a combined analysis of 29 cities there was evidence of a systematic change in various part 1 crime rates. We also are the first study to consider the external validity of the 29 cities included relative to all 70 largest cities in the US. Furthermore, combined analyses of cities have not used data beyond the summer of 2020, which prevents these analyses from fully examining seasonal trends or the persistence of any crime rate changes. Second, we assembled data on the exact timing and strictness of the lockdown over time for each city during the lockdown period which allows us to include an index of openness in our analysis. Third, we focus specifically on the impact of the lockdown and subsequent opening up on crime throughout the year whereas other analyses (eg., \citep{coote_fbi_2020}) with the exception of \citep{rosenfeld_pandemic_2020}, focus more on the question of whether homicide in particular overall increased in 2020 including the pre-lockdown months. 

Because we have data available from after initial lockdowns were relaxed throughout the summer and fall and into early winter, thus allowing seasonal trends to be properly detected, we further contribute to previous work by illustrating how evolving policies surrounding the pandemic were associated with possible changes in crime rates. Using rigorous statistical models, we are able to account for seasonality as well as year-to-year changes in homicide and other crime trends across many U.S. cities. Furthermore, we are also unaware of any articles besides work by \citep{rosenfeld_pandemic_2020} based on an entirely different methodology that adapts a time series method to panel data, that has analyzed crime trends following the killing of George Floyd, which we are able to do due to our 2020 data coverage continuing throughout the entire year.

\section{Data and Methods}
\label{sec:datamethods}
\subsection{Data Sources}

The outcome variables of interest are per capita crime rates by type of crime. We used publicly available crime incident report data from 29 of the 70 largest cities in the United States \citep{worldpopulationreviewcom_200_2020}. These 29 cities are those that made daily- or weekly-level data publicly available for the full years of 2018 and 2019, and at least the first six months of 2020 in an accessible and usable format. For example, four cities (Nashville, Louisville, Houston, and Seattle) in our data set published NIBRS data, which was straightforward to categorize into crime types. In the appendix directory of our github repository\footnote{All of the data and scripts used for this paper can be found in the covid-crime repository: \url{https://github.com/mmeyer717/covid-crime}}, a CSV file (``appendix-table-a1.csv'') shows which variable in each city’s data file was used to identify crime descriptions and which crime descriptions or codes for each city were classified into each of these five Part 1 crimes. To determine rates per capita, we used U.S. Census Bureau Subcounty Resident Population Estimates from 2010-2019 \citep{us_census_bureau_city_2020}. 

To assess the external validity of our study, in Appendix C, Figures \ref{fig:hom-ev}-\ref{fig:larceny-ev} show the crime data in 2018 for each of the five crime types we analyze for the 29 cities in our analysis compared with all of the 70 most populous U.S. cities. We compare the distribution of yearly city crime rates per 100,000 residents for our data (after aggregating to annual rates), relative to UCR annual data for all of the top 70 largest cities. Based on these figures, we conclude that the distribution of yearly crime rates was similar for the 29 cities included in this study and all 70 largest cities in 2018. The one exception is homicide where the interquartile range for the 29 cities in our data is wider than that of all the top 70 most populous cities. This suggests that the 29 cities cover the range of homicide rates that occur in the top 70 largest cities and are somewhat more variable. In summary, we find that the data from our non-random sample of the top 70 most populous cities is quite representative of this population based on 2018 crime rates. 

We identified data on five of seven FBI UCR part 1 crimes: homicide, robbery, burglary, larceny, and auto theft. The two Part 1 crime types we do not include are rape and aggravated assault. Though 26 of these cities made data about crime reports related to rape available, due to the inconsistent nature of classifying a crime as rape across cities, we decided to not include rape in our analyses. For aggravated assault, data was widely available from these cities, but the 2018 aggregated yearly totals from the weekly data were on average 20\% lower than the 2018 UCR reported yearly totals, which caused us to be concerned about the quality of the weekly aggravated assault count, largely due to inconsistency about classifying crimes as ``aggravated' versus less serious assault. Because rape and aggravated assault are both serious forms of violence their omission from the analysis is unfortunate.  We thus join with other researchers and policy makers in calling for more consistent records of these two important crime types. Appendix A, Table \ref{table:a2} lists the 29 cities and which analyses we included them in. 

We also gathered data from media sources \citep{cnn_this_2020, the_new_york_times_see_2020} and state and county health departments regarding lockdown dates for all 29 cities and the timing and nature of pandemic-related lockdown reopening phases during the spring and summer of 2020 for 24 of the 29 cities. From this information we created two different measures of the lockdown stage a given city was in during a particular week. The first measured in two-week intervals the amount of time that has passed since the initial lockdown measure was put in place in each city. These do not align with calendar time as different cities began their lockdowns at different times. The other was a measure of the \textit{extent} to which each city was open or locked down in each week which we refer to as an ``openness index''. The index was on a scale from 0 (least open) to 14 (most open). For all weeks leading up to a city’s COVID-related lockdown week, this index took a value of 14. The value for the index for the weeks after a city’s lockdown were determined by how restricted certain sectors of the economy were. The sectors we considered were ``Entertainment'', ``Food and Drink'', ``Industries'', ``Outdoor and Recreation'', ``Personal Care'', ``Places of Worship'', and ``Retail,'' and each sector could be assigned a value of 0 (no components of this sector were operating at any capacity), 1 (at least some industries in the sector were not operating at full capacity), or 2 (all industries in the sector were operating at full capacity). We visualize the variation in the openness index in 2020 across all available cities in Appendix A, Figure \ref{fig:open}. For more information about what industries fell under each sector, please refer to Appendix A, Table \ref{tabel:a3}.

\subsection{Method}

To measure the associations between COVID-19 pandemic-related lockdowns and crime rates, we use two different sets of linear regression models. Our unit of analysis is a city-week pair in all models. One set of models tests whether seasonal crime trends in 2020 are significantly different from those in the past two years. This model includes data for all of 2020, encompassing months before the pandemic lockdowns, the spring lockdown and re-opening phases, summer mass protests sparked by the killing of George Floyd by police in Minneapolis, and the remainder of 2020 following any major protest activity as the pandemic continued. The other set of models focuses on describing crime patterns in relation to specific pandemic-related shutdown timing and extent in the first couple of months following the pandemic compared to crime trends from the past two years. These models include 2020 data from January through week 21, the week of the killing of George Floyd and the week before ensuing protests occurred in most cities across the U.S. In all of our models, we cluster standard errors at the city-level.

\subsubsection{Response Variable}
For both sets of models, we used the same response variables. Let $i = 1,\ldots, 5$ represent the different types of crime we analyze, and let $c = 1,\ldots, 29$, $r = 2018, 2019, 2020$, and $w = 1, \ldots, 52$ index the city, year, and week of the year of interest respectively. For each crime type, $i=1,\ldots,5$, we calculate the response variable as:
\begin{equation}
y_{cwr}= \text{crime rate}_{cwr}*52-\text{annual crime rate}_{c}
\end{equation}
where the weekly city crime rate for each crime is the number of incidents per 100,000 residents, and the annual crime rate for each city and crime is the total number of incidents in 2018 per 100,000 residents. In words, our response variable is the difference between what would be the annual crime rate for that city and crime, assuming that the crime rate in that particular week held for all 52 weeks of the year, and the annual crime rate in 2018. Differencing from the 2018 city rate is an alternative to including city fixed effects in the model, both specifications give similar results.\footnote{Results for the model specification with city fixed effects are available upon request.}

\subsubsection{Model of Seasonal Crime Rates in 2020}
Our first set of models explores how 2020 seasonal crime trends differed from seasonal crime trends in 2018 and 2019 using month indicators. For each crime type, $i = 1, ..., 5$, we modeled the weekly response variable as follows:

\begin{equation}
\begin{split}
    y_{cwr} = & \beta_0 + \beta_1*\text{Year2019}_{r} + \beta_2*\text{Month}_{w} + \beta_3*\text{Pre-Pandemic 2020}_{wr} + \\
    & \beta_4*\text{Lockdown 2020}_{wr} + \beta_5*\text{Summer Protests 2020}_{wr} + \\
    & \beta_6*\text{End of Year 2020}_{wr} + \epsilon_{cwr} 
\end{split}
\end{equation}

We estimate a fixed effect for the year 2019 relative to the omitted year of 2018, and we also estimate fixed effects for the months February through December relative to the omitted month of January. For 2020, we classify January and February as ``Pre-Pandemic'', March through May as ``Lockdown'', June through August as ``Summer Protests'', and September through December as ``End of Year''; we estimate fixed effects for each of these time periods which thereby precludes the need for a 2020 fixed effect. We further analyze how crime rates changed in the different periods of 2020 by conducting post-estimation tests with the coefficients and standard errors from this model (Eq. (2)). We conduct two-sided t-tests at the 5\% level to determine whether the coefficients for the later periods of 2020 are significantly different from the coefficient for the pre-pandemic period.

Week units are seven days long, and we prioritized keeping approximated month units to be composed of these full week units. This results in approximate month units consisting of four or five full weeks as seen in Table \ref{table:a4} in the Appendix. With the exception of Lincoln, Nebraska, whose late 2020 data we could not find, and San Francisco which did not report weekly homicide counts, we included all other cities in this analysis.

\subsubsection{Model of Crime Rates by Weeks into Lockdown}
To understand how relationships between crime rates and pandemic-related lockdowns may vary over the duration of the lockdowns in, we used two different types of models. For both types of models, we restricted the data to end in week 21 of 2020, since the protests following George Floyd’s killing might have impacted crime rates differently than the lockdowns, and many cities had significantly loosened lockdown restrictions by this time.
One type modeled the response variable as such, for each crime type, $i = 1,\ldots,5$: 
\begin{equation}
\begin{aligned}
    y_{cwr} &= \beta_0 + \beta_1*\text{Year}_{r} + \beta_2*\text{Month}_{w} + \\
    & \sum_{(j,k) \in {(-1,0),(1,2),...,(9,10)}} \beta_{(j,k)}*\text{Weeks since Lockdown}_{(j,k)_{cwr}} + \epsilon_{cwr}
\end{aligned}
\end{equation}

Similar to the model described in Eq. (2), the intercept term reflects crime rates in January 2018. We again estimated fixed effects for 2019 and 2020 and the months February through December. 
We now also include biweekly lockdown fixed effects for the number of weeks since city c’s lockdown began. To illustrate, we estimate one fixed effect across all cities for one week before and the week of a city’s lockdown initiation, another fixed effect across all cities for the first and second full weeks of a city’s lockdown, and so forth. Because cities enacted lockdown measures during different weeks, these fixed effects do not match up with two particular calendar weeks. 

\subsubsection{Model of Crime Rates by Openness Index}

The second type of model we used to analyze how the evolution of pandemic-related lockdowns over time in 2020 might have been associated with a change in crime rates modeled the response variable as described below, for each crime type, $i = 1,\ldots, 5$:
\begin{equation}
    y_{cwr} = \beta_0 + \beta_1*\text{Year}_{r} + \beta_2*\text{Month}_{w} + \beta_3*\text{Openness Index}_{cwr} + \epsilon_{cwr}
\end{equation}

Again, our baseline is January 2018. Our model is the same as Eq. (3), except instead of estimating biweekly lockdown fixed effects for the number of weeks since city $c$’s lockdown, we look at the association between crime rates and the weekly openness index developed as discussed in the data section.

\section{Results}
Our discussion of results is organized as follows. Results are presented by crime type with three analyses discussed for each: overall 2020 differences from typical seasonal trends (Eq. (2)), more granular 2020 differences specifically during the pandemic lockdown in spring 2020 (Eq. (3)), and whether differences during the pandemic lockdown were associated with degree of city openness (Eq. (4))\footnote{As mentioned in Section 2.2.3, the response variable for this model is measured at the weekly-level despite using biweekly indicators.}. We first discuss findings for the two crime types, homicide and auto theft, for which we find statistically significant increases in two or more of the 2020 post-pandemic periods compared to the average seasonal trends across 2018 and 2019. Next, we discuss the results for burglary for which we found no sustained statistically significant differences in 2020 post-pandemic periods compared with average prior seasonal trends. We conclude with a discussion of our results for robbery and larceny for which we found statistically significant decreases during 2020. Both robbery and larceny rates decreased relative to prior years during the spring lockdown of 2020 and these declines persisted throughout the summer protest and end of the year periods. As described below, larceny was the only crime type for which changes in the openness index across cities and weeks was associated with statistically significant weekly changes in crime rates.

\subsection{Homicide}

\begin{figure}[h]
    \centering
    \includegraphics[scale = 0.6]{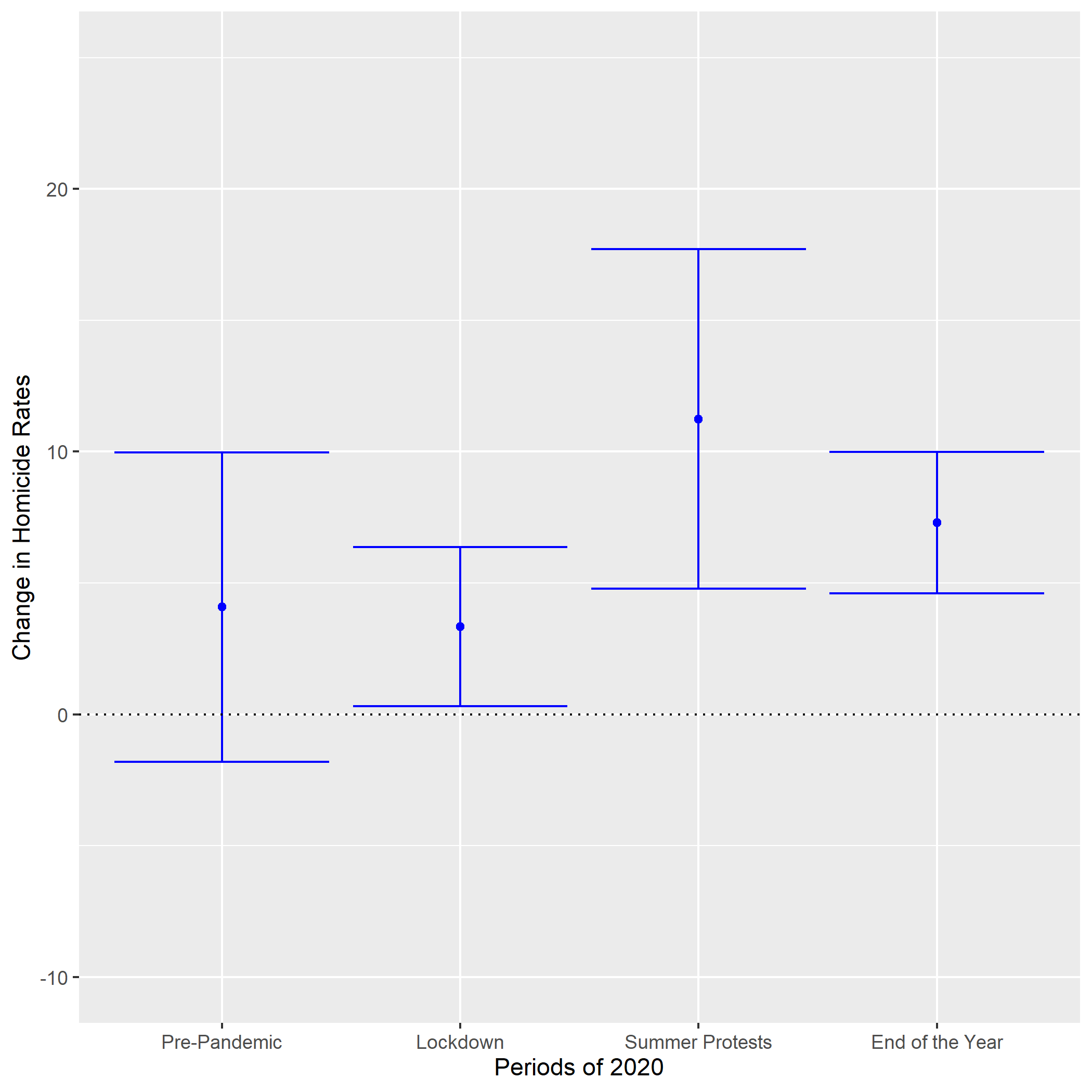}
    \caption{Changes in Homicide Rates in 2020 Compared to Typical Seasonal Trends with 95\% Confidence Intervals}
    \label{fig:hom_month}
\end{figure}

Figure \ref{fig:hom_month} summarizes our analysis of homicide rates in 2020 based on the seasonal adjustment regression that divides 2020 into four distinct pre- and post-pandemic periods specified by Eq. (2). The black dashed horizontal zero line represents the expected homicide rate based on seasonal trends in the prior two years. The solid blue dots indicate the average increase in homicide rates in 2020 relative to seasonal trends in the prior two years. The vertical lines are 95\% confidence intervals for the 2020 effect estimates. Unless otherwise stated, all hypothesis tests are at the .05 level for a two-tailed test. The coefficient table for this model (Table \ref{table:hom-month}), the post-estimation t-tests for the difference between later time periods and the pre-pandemic period (Table \ref{table:hom-postest}), and the figure of monthly seasonal trends in homicide rates in the prior two years plus 2020 changes relative to these trends (Figure \ref{fig:hom_month_oldfig1}) can be found in Appendix B.  

As seen in Figure \ref{fig:hom_month}, both before the pandemic began and during the spring pandemic lockdown, homicide rates across the cities in this analysis were about 4 homicides per 100,000 people higher than in prior years. For reference, the average homicide rate for these 29 cities in 2018-19 was 16.4 homicides per 100,000 people. This difference does not reach statistical significance during the pre-pandemic period but does during the spring lockdown. During the summer protest period of 2020, the homicide rate rose to more than 11 homicides per 100,000 people higher than usual summer levels ($p < 0.001$). This also was significantly higher than it had been during the spring lockdown period ($p < 0.01$). It remained above spring lockdown levels ($p < 0.05$) at about 7 additional homicides per 100,000 people for the rest of 2020 after protests ended but the pandemic continued. 

\begin{figure}[h]
    \centering
    \includegraphics[scale = 0.75]{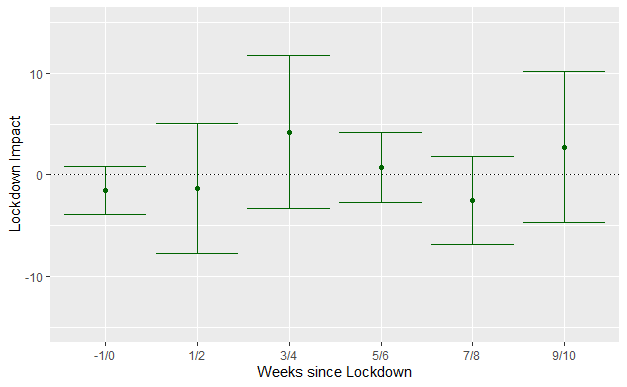}
    \caption{Biweekly 2020 Differences in Homicide Rates during the Lockdown relative to Typical Seasonal levels with 95\% confidence intervals}
    \label{fig:hom_biweek}
\end{figure} 

Next, we focus in greater detail on the initial pandemic lockdown period of 2020 in particular. We report analyses examining how the number of weeks into lockdown might have been associated with a change in weekly homicide rates during the spring of 2020. The estimated biweekly indicator variables’ coefficients are displayed in Figure \ref{fig:hom_biweek} along with 95\% confidence intervals. We find that none of these coefficients are statistically significantly different from 0, which suggests the weekly homicide rate did not on average change right before or in the weeks following lockdowns being put in place relative to typical seasonal trends, despite the overall increase in this period found in the prior model. Finally, we find that changes in the openness index across this time period and across cities were not statistically significantly associated with changes in the weekly homicide rate (p-value = 0.35). The coefficient tables for both of these models can be found in the appendix (Appendix B, Tables \ref{table:hom-biweek} and \ref{table:hom-open}). Thus, our analysis found evidence of sustained increases in the homicide rate across the cities included in our analysis that began during the spring lockdown and grew further in the summer protest period and the remainder of 2020.

\subsection{Auto Theft}
\begin{figure}
    \centering
    \includegraphics[scale=0.6]{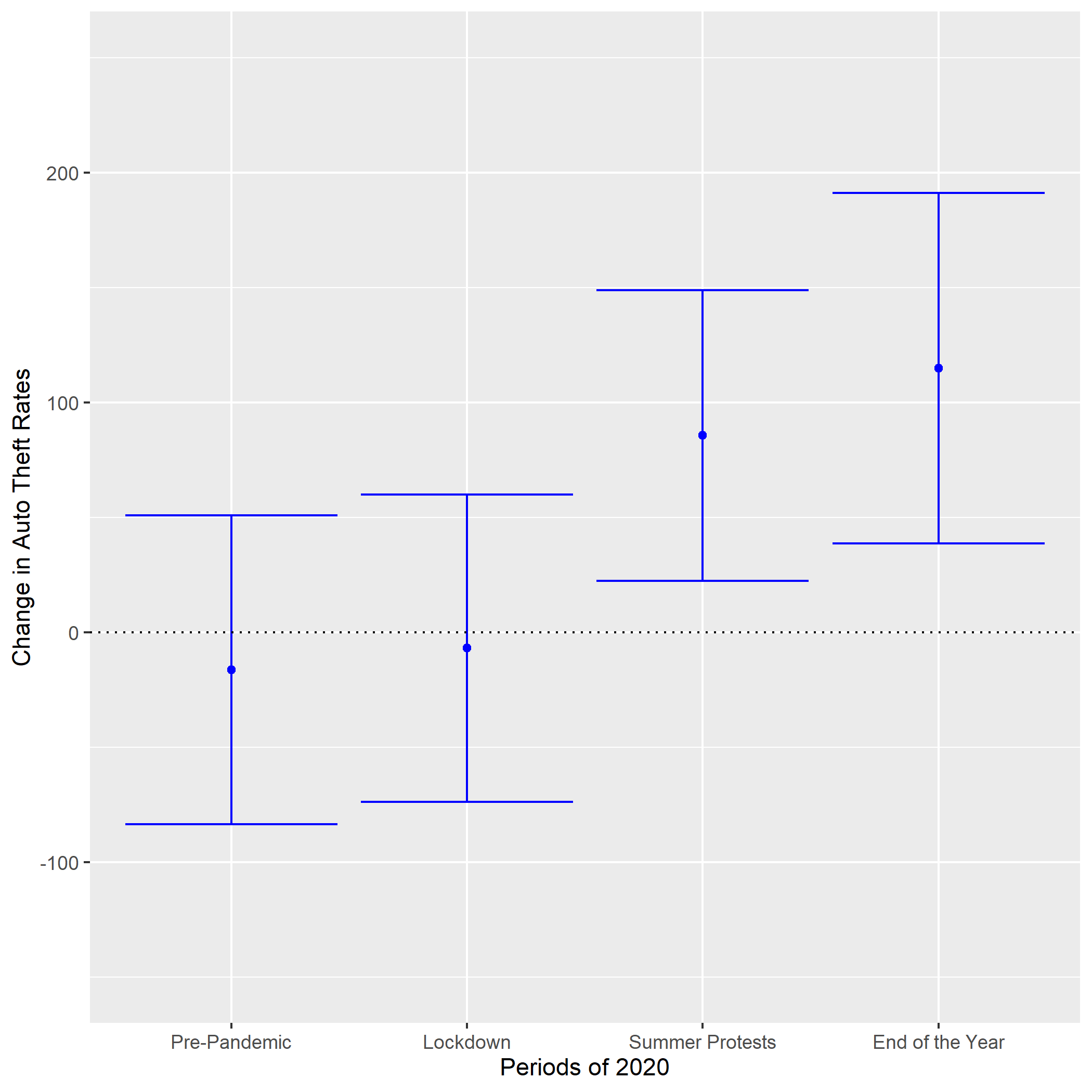}
    \caption{Changes in Auto Theft Rates in 2020 Compared to Typical Seasonal Trends with 95\% Confidence Intervals}
    \label{fig:auto_month}
\end{figure}

Figure \ref{fig:auto_month} is the counterpart of Figure \ref{fig:hom_month} for auto theft. As seen in Figure \ref{fig:auto_month}, in the first months of 2020 before the pandemic began, auto theft rates were similar to those in the same months of the prior two years, and they remained similar to prior trends during the spring pandemic lockdown period. For reference, the average auto theft rate for these 29 cities in 2018-19 was 534 auto thefts per 100,000 people. Beginning in the summer protest period, auto theft rates increased by 85 on average relative to expected summer levels ($p < 0.01$) and relative to pre-pandemic 2020 rates ($p < 0.01$). They then remained high, 115 more auto thefts per 100,000 population, throughout the rest of 2020 relative to usual fall and winter rates and relative to pre-pandemic 2020 rates ($p < 0.01$ for both).  

\begin{figure}
    \centering
    \includegraphics[scale=0.75]{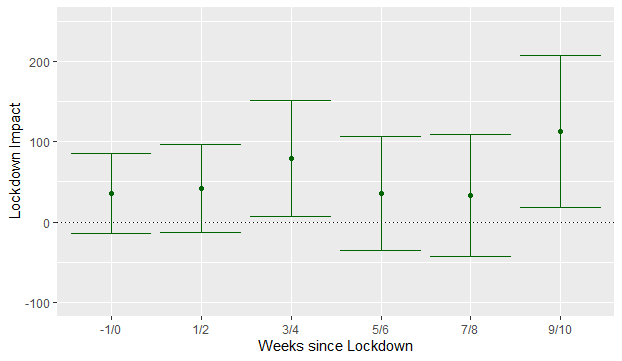}
    \caption{Biweekly 2020 Differences in Auto Theft Rates during the Lockdown relative to Typical Seasonal levels with 95\% confidence intervals}
    \label{fig:auto_biweek}
\end{figure}

Furthermore, when we look more closely at the spring lockdown period, we find that there was a statistically significant increase in weekly auto theft rates at the very end of the lockdowns - in weeks 9-10 after each city enacted their own lockdown measures ($p < 0.05$). As seen in Figure \ref{fig:auto_biweek}, the change for all other lockdown weeks was not statistically significant at the .05 level but lockdown weeks 3-4 show an increase at the 0.10 level ($p = 0.059$). Similar to homicide, we find that changes in the openness index over the lockdown time period and over cities were not associated with changes in the weekly auto theft rate ($p = 0.46$). Model results for auto theft rates and post-estimation t-tests results for the comparison of later 2020 time periods to the pre-pandemic period can be found in Appendix B (Tables \ref{table:auto-month}, \ref{table:auto-postest}, \ref{table:auto-biweek}, and \ref{table:auto-open}).

\subsection{Burglary}

\begin{figure}
    \centering
    \includegraphics[scale=0.6]{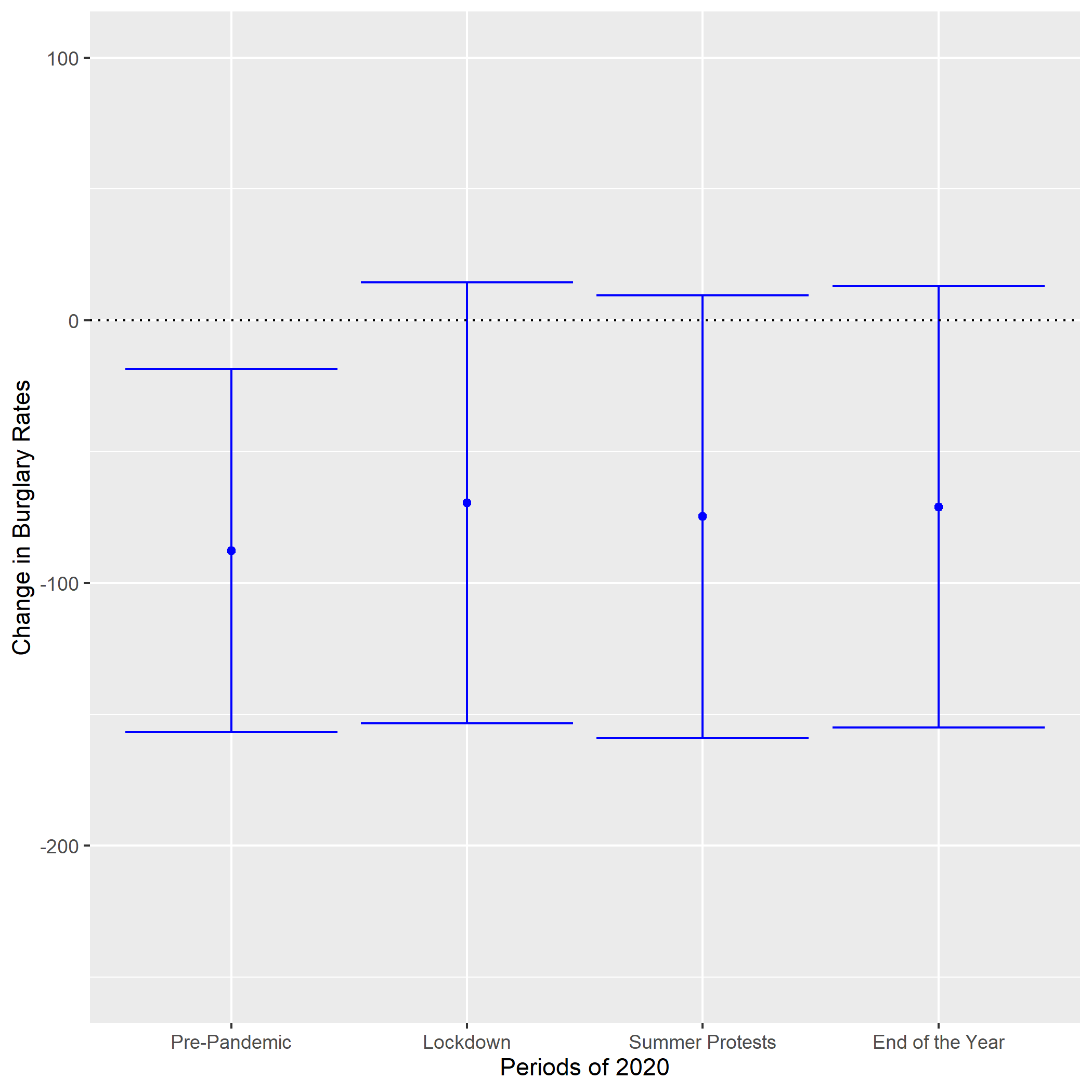}
    \caption{Changes in Burglary Rates in 2020 Compared to Typical Seasonal Trends with 95\% Confidence Intervals}
    \label{fig:burg_month}
\end{figure}

As seen in Figure \ref{fig:burg_month}, in the first months of 2020 before the pandemic began, burglary rates were lower, by 88 burglaries per 100,000, than in the first months of the prior two years ($p < 0.05$). For reference, the average burglary rate for these 29 cities in 2018-19 was 623 burglaries per 100,000 people. While the burglary rates remained low during the spring pandemic lockdown, summer protest, and end of the year periods these differences relative to expected seasonal rates were not statistically significant at the 0.05 level (summer and end of 2020 differences were marginally significant at the 0.10 level). We also do not detect differences in the 2020 burglary rates (relative to seasonal trends) between these four periods. 

\begin{figure}
    \centering
    \includegraphics[scale=0.75]{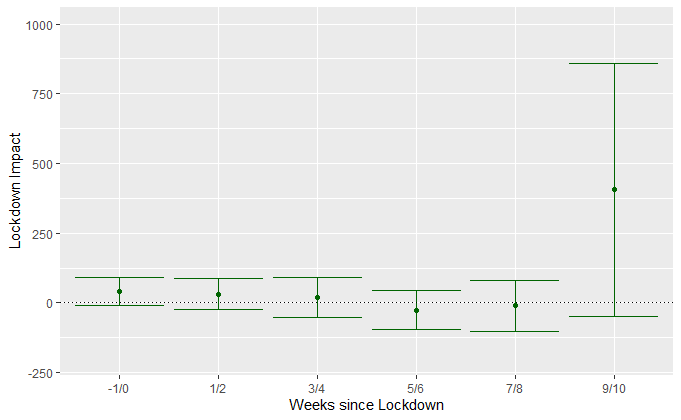}
    \caption{Biweekly 2020 Differences in Burglary Rates during the Lockdown relative to Typical Seasonal levels with 95\% confidence intervals.}
    \label{fig:burg_biweek}
\end{figure}

As seen in Figure 6, we further find that the burglary rate did not significantly change right before or in the weeks following lockdowns being put in place relative to typical seasonal trends. Note that the large standard error and estimate for weeks 9 and 10 after lockdown are likely due to increased heterogeneity across cities in particular due to Saint Paul, MN experiencing a large increase in burglaries in week 21 and also to there being fewer observations (24) available for the estimation of this effect depending on when cities began their lockdowns. To test how sensitive our results are to this outlier, we also re-ran this model disregarding week 21 in St. Paul. As seen in Appendix B, 
Figure \ref{fig:burg_no_stp} and Table \ref{table:burg-biweek-nostp}, weeks 9 and 10 after lockdowns saw statistically significantly higher burglary rates compared to typical seasonal trends when this observation is omitted. We also find that changes in the openness index over the lockdown time period and over cities are not significantly associated with changes in the weekly burglary rate ($p = 0.18$). All of the coefficient tables for the models of the burglary rate and post-estimation t-tests results for the comparison of later 2020 time periods to the pre-pandemic period can be found in (Appendix B, Tables \ref{table:burg-month}, \ref{table:burg-postest}, \ref{table:burg-biweek}, and \ref{table:burg-open}). 

\subsection{Robbery}
\begin{figure}
    \centering
    \includegraphics[scale=0.6]{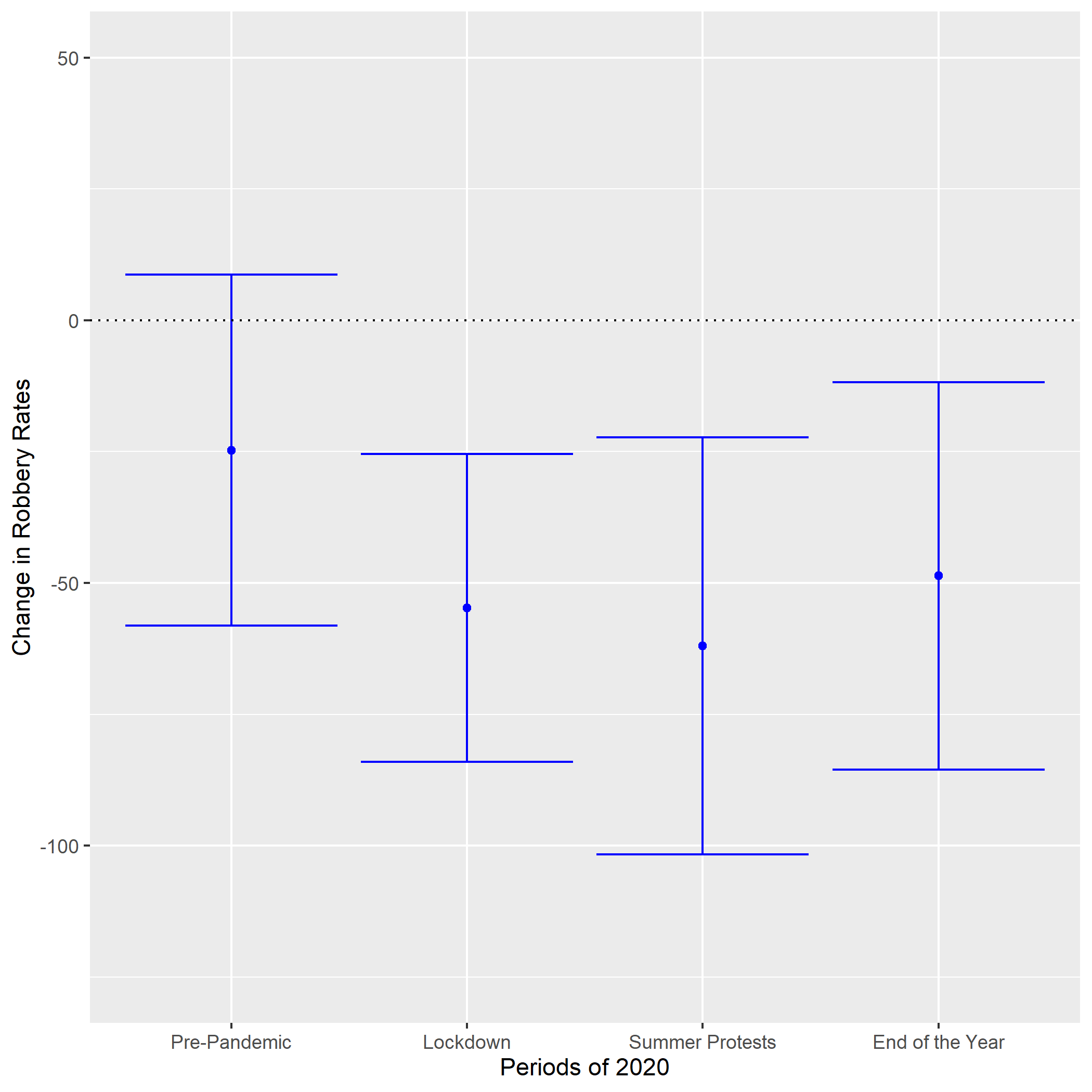}
    \caption{Changes in Robbery Rates in 2020 Compared to Typical Seasonal Trends with 95\% Confidence Intervals}
    \label{fig:rob_month}
\end{figure}
As seen in Figure \ref{fig:rob_month}, in the first two months of 2020 before the pandemic began, robbery rates across the cities in this analysis were somewhat lower but not statistically significantly different on average than in the same months of prior years. For reference, the average robbery rate for these 29 cities in 2018-19 was 287 robberies per 100,000 people. During the spring pandemic lockdown, the robbery rate decreased substantially by 55 robberies per 100,000 people relative to expected seasonal rates ($p < 0.001$) and also decreased relative to pre-pandemic rates ($p < 0.01$). During the summer protest period and in the rest of 2020 the robbery rate remained similarly low ($p < 0.01$). Seasonally adjusted robbery rates during the spring lockdown and summer protest periods are statistically significantly lower than pre-pandemic period rates ($p < 0.05$ for both) but end of year seasonally adjusted rates do not significantly differ.

\begin{figure}
    \centering
    \includegraphics[scale=0.75]{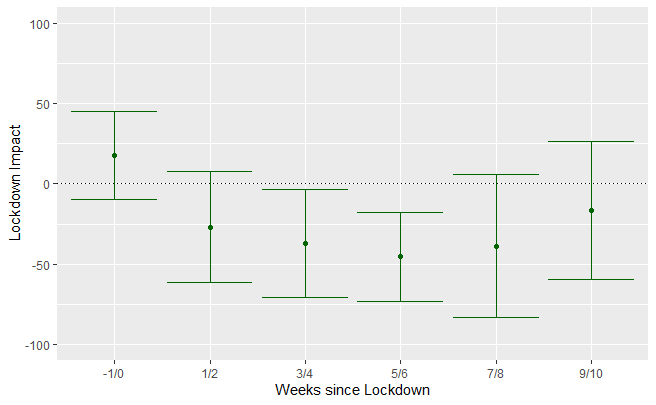}
    \caption{Biweekly 2020 Differences in Robbery Rates during the Lockdown relative to Typical Seasonal levels with 95\% confidence intervals}
    \label{fig:rob_biweek}
\end{figure}

Figure \ref{fig:rob_biweek} reports the results of the model focusing in on robbery rates during the lockdown. Though the weeks immediately before and after lockdowns went into place are not associated with statistically significant changes in weekly robbery rates, weeks 3 - 4 and weeks 5- 6 are both associated with statistically significant decreases relative to usual levels at these times of year. However, these decreases were not sustained in weeks 7-10. There is evidence at the .10 level ($p = 0.09$) that robbery rates were higher during the lockdown in cities and weeks that were more open relative to those that were more closed or more locked-down. The coefficient tables for the models analyzing robbery rates and post-estimation t-tests results for the comparison of later 2020 time periods to the pre-pandemic period can be found in Appendix B (Tables \ref{table:rob-month}, \ref{table:rob-postest}, \ref{table:rob-biweek}, and \ref{table:rob-open}). 

\subsection{Larceny}
\begin{figure}
    \centering
    \includegraphics[scale=0.6]{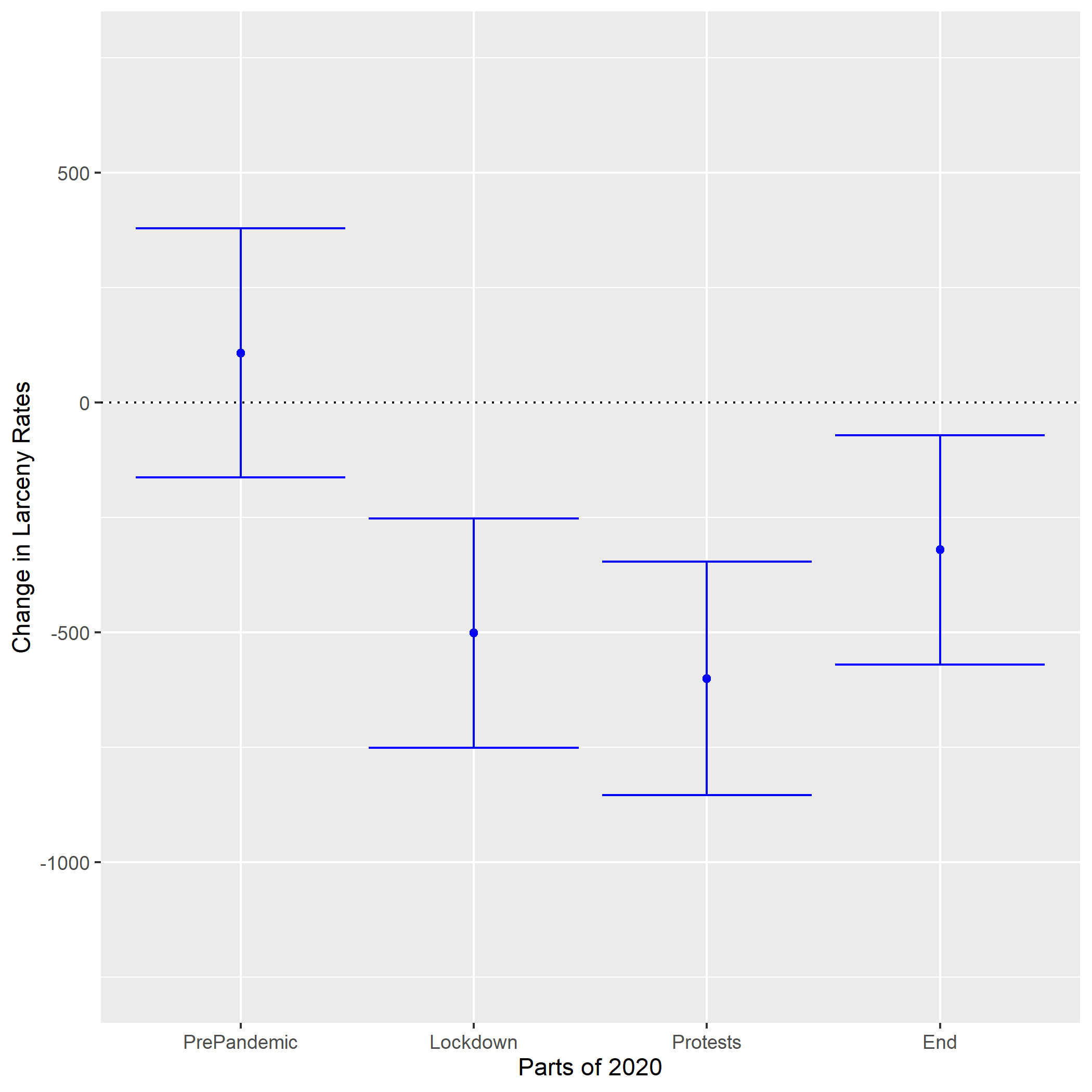}
    \caption{Changes in Larceny Rates in 2020 Compared to Typical Seasonal Trends with 95\% Confidence Intervals}
    \label{fig:larceny_month}
\end{figure}
Figure \ref{fig:larceny_month} shows that during the 2020 pre-pandemic period, larceny rates were similar to those in the same months of prior years. For reference, the average larceny rate for these 29 cities in 2018-19 was 2,642 larcenies per 100,000 people. During the spring pandemic lockdown, the larceny rate decreased dramatically, by 502 larcenies per 100,000 people relative to expected seasonal rates and decreased relative to pre-pandemic 2020 rates ($p < 0.001$ for both). During the summer protest period, larceny rates remained similarly low, 600 fewer larcenies per 100,000 people compared to expected seasonal levels ($p < 0.001$). A decrease in larceny rates relative to expected seasonal levels persisted for the rest of 2020 ($p < 0.05$ relative to seasonal trends, $p < 0.001$ relative to 2020 pre-pandemic rates).  

\begin{figure}
    \centering
    \includegraphics[scale=0.75]{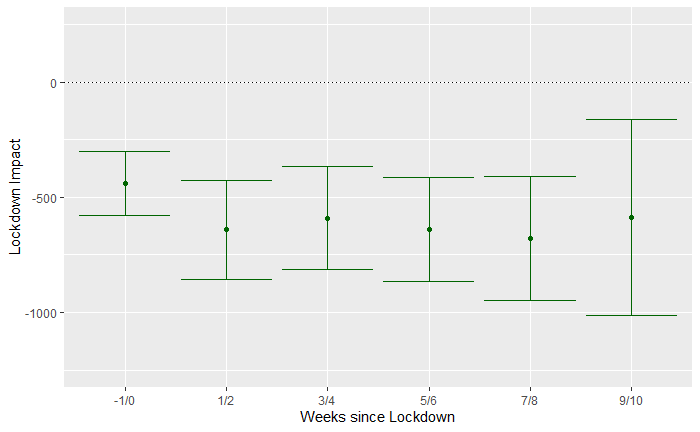}
    \caption{Biweekly 2020 differences in larceny rates during the lockdown relative to typical seasonal levels with 95\% confidence intervals}
    \label{fig:larceny_biweek}
\end{figure}

Figure \ref{fig:larceny_biweek} reports the results of the model focusing in on larceny rates during the lockdown. Here again we find evidence of a sustained decline that is even more consistent than for robbery. Evidence of this decline is also supported by analyses of the association of the openness index with larceny rates in which we find a positive and statistically significant association (p < 0.001) which implies that when each city’s specific lockdown restrictions were eased (cities were relatively more open) larceny rates increased.  The coefficient tables for the models analyzing weekly larceny rates and post-estimation t-tests results for the comparison of later 2020 time periods to the pre-pandemic period can be found in Appendix B, Tables \ref{table:larceny-month}, \ref{table:larceny-postest}, \ref{table:larceny-biweek}, and \ref{table:larceny-open}. 

\section{Discussion and Conclusions}
We find that changes in crime rates during the pandemic from March through December 2020 compared to counterpart periods in 2018 and 2019 were mixed. For homicide we find significant increases through all three pandemic periods (lockdown, summer protests, and end of the year) that are not, however, significantly different from a non-significant increase in the pre-pandemic period of 2020. We also find significant increases in auto theft during the final half of 2020 spanned by the summer protest period and the end of year period. These increases are also significant compared to the 2020 pre-pandemic period. Balanced against these increases were declines in robbery and larceny throughout the pandemic and some indication of declines in burglary in which all pandemic period point estimates are negative but not statistically significant at the 5\% level. The declines in robbery and larceny during the pandemic periods were also significant relative to the 2020 pre-pandemic period. We generally did not find associations between the openness index we developed and crime rates suggesting that, except for larceny, incremental changes in cities lockdown policies over time and across cities were not associated with detectible changes in crime rates. 

Our findings overall are a puzzle for future research to solve. Crime rates by type (e.g., homicide and robbery) usually move in tandem but this did not occur in the pandemic period studied. We instead found a sustained increase in homicide throughout the pandemic and in auto thefts beginning in June and sustained declines in larceny and robbery throughout the pandemic periods. None of the pandemic period estimates for burglary were significant at the 0.05 level.  As we noted in the introduction our aim was to document changes in crime rates during the pandemic and initial lockdown period but not to ascertain their cause. Regarding the causes of the patterns revealed by our analysis, we can only speculate. One admittedly post-hoc and only partial explanation for this unusual pattern relates to opportunity. The decline in robbery rates may have been caused by reductions in numbers of people on streets, particularly at night, thereby reducing robbery targets. Larceny rates may have declined due to reduced foot traffic and possibly increased security (for mask enforcement) in brick-and-mortar retail outlets. The increase in auto thefts may have resulted from the combination of more vehicles sitting idle on the streets and fewer people outside, including police, making it easier to surreptitiously steal the vehicles. Opportunity, however, does not easily explain the findings for homicide and burglary. More time sheltering at home coupled with increased drug and alcohol intake might have given rise to more conflicts that end in a domestic homicide, but press reports on the homicide increase do not attribute it to increases in domestic homicides. Concerning non-domestic homicides, reduced traffic in places such as bars and reductions in young men loitering on the streets due to pandemic restrictions should lead to reductions, not increases, in the types of conflicts that give rise to homicides.  As for burglary, more time sheltering at home would reduce opportunities for domestic burglaries, consistent with the trend in the data here. On the other hand, closed businesses might increase opportunities for non-residential burglaries. Perhaps for burglary, these opposing forces balanced out to no systematic detectable change, but that reasoning may be a stretch. 

Our bottom line on policy recommendations is that in the absence of a clear understanding of how the pandemic affected crime and in the spirit of evidence-based crime policy, we caution against advancing policy at this time based on lessons learned from the pandemic “natural experiment”. The pandemic natural experiment does, however, provide a unique opportunity for testing various theories of the causes of crime, particularly those related to opportunity which includes police presence and contemporaneous social and economic stressors (e.g., poverty, substance abuse, and depression). We recommend that future research examine crime changes during the pandemic based on more complete crime data than used here. Our study like all others done to date have been based on analyses of cities for which data was available as the pandemic was unfolding. In all studies conducted thus far the cities included were far from a complete census of all cities or a random sample of all cities. Further, to our knowledge there have been no studies of impacts outside of cities in suburban and rural areas or in later stages of the pandemic in 2021. Such studies should be a priority.

\section{Acknowledgements}
This work was supported by the National Science Foundation (NSF) under RAPID Grant Agreement No. 2029890. Mikaela Meyer was supported by the National Science Foundation Graduate Research Fellowship Program under Grant No. DGE1745016.

\bibliographystyle{unsrtnat}
\bibliography{bibliography}

\begin{thebibliography}{19}
\providecommand{\natexlab}[1]{#1}
\providecommand{\url}[1]{\texttt{#1}}
\expandafter\ifx\csname urlstyle\endcsname\relax
  \providecommand{\doi}[1]{doi: #1}\else
  \providecommand{\doi}{doi: \begingroup \urlstyle{rm}\Url}\fi

\bibitem[Asher and Horwitz(2020)]{asher_its_2020}
Jeff Asher and Ben Horwitz.
\newblock It’s {Been} ‘{Such} a {Weird} {Year}.’ {That}’s {Also}
  {Reflected} in {Crime} {Statistics}. - {The} {New} {York} {Times}, July 2020.
\newblock URL
  \url{https://www.nytimes.com/2020/07/06/upshot/murders-rising-crime-coronavirus.html}.

\bibitem[Hilsenrath(2020)]{hilsenrath_homicide_2020}
Jon Hilsenrath.
\newblock Homicide {Spike} {Hits} {Most} {Large} {U}.{S}. {Cities}.
\newblock \emph{Wall Street Journal}, August 2020.
\newblock ISSN 0099-9660.
\newblock URL
  \url{https://www.wsj.com/articles/homicide-spike-cities-chicago-newyork-detroit-us-crime-police-lockdown-coronavirus-protests-11596395181}.

\bibitem[McCarthy(2020)]{mccarthy_major_2020}
Niall McCarthy.
\newblock Major {American} {Cities} {See} {Sharp} {Spike} {In} {Murders} {In}
  2020 [{Infographic}], August 2020.
\newblock URL
  \url{https://www.forbes.com/sites/niallmccarthy/2020/08/04/major-american-cities-see-sharp-spike-in-murders-in-2020-infographic/}.
\newblock Section: Business.

\bibitem[Struett(2020)]{struett_chicago_2020}
David Struett.
\newblock Chicago nears 700 homicides in 2020, a milestone reached just one
  other time since 1998, November 2020.
\newblock URL
  \url{https://chicago.suntimes.com/crime/2020/11/18/21573378/chicago-homicides-700-murders-2020-gun-violence-shootings}.

\bibitem[Campbell(2020)]{campbell_violent_2020}
Josh Campbell.
\newblock Violent crime soars during pandemic as confidence in police takes a
  hit - {CNN}, August 2020.
\newblock URL
  \url{https://www.cnn.com/2020/08/16/us/violent-crime-soars-confidence-in-police-takes-hit/index.html}.

\bibitem[Becker and Corley(2020)]{becker_violent_2020}
Deborah Becker and Cheryl Corley.
\newblock Violent {Crime} {Increases} {In} {Several} {Cities} {Nationwide},
  August 2020.
\newblock URL
  \url{https://www.npr.org/2020/08/14/902456117/violent-crime-increases-in-several-cities-nationwide}.

\bibitem[FBI(2020)]{fbi_overview_2020}
FBI.
\newblock Overview of {Preliminary} {Uniform} {Crime} {Report},
  {January}–{June}, 2020 — {FBI}, 2020.
\newblock URL
  \url{https://www.fbi.gov/news/pressrel/press-releases/overview-of-preliminary-uniform-crime-report-january-june-2020}.

\bibitem[Lopez(2020)]{lopez_rise_2020}
German Lopez.
\newblock The rise in murders in the {US}, explained, August 2020.
\newblock URL
  \url{https://www.vox.com/2020/8/3/21334149/murders-crime-shootings-protests-riots-trump-biden}.

\bibitem[Piquero et~al.(2020)Piquero, Riddell, Bishopp, Narvey, Reid, and
  Piquero]{piquero_staying_2020}
Alex~R. Piquero, Jordan~R. Riddell, Stephen~A. Bishopp, Chelsey Narvey, Joan~A.
  Reid, and Nicole~Leeper Piquero.
\newblock Staying {Home}, {Staying} {Safe}? {A} {Short}-{Term} {Analysis} of
  {COVID}-19 on {Dallas} {Domestic} {Violence}.
\newblock \emph{American Journal of Criminal Justice}, pages 1--35, June 2020.
\newblock ISSN 1066-2316.
\newblock \doi{10.1007/s12103-020-09531-7}.
\newblock URL \url{https://www.ncbi.nlm.nih.gov/pmc/articles/PMC7293590/}.

\bibitem[Campedelli et~al.(2020)Campedelli, Aziani, and
  Favarin]{campedelli_exploring_2020}
Gian~Maria Campedelli, Alberto Aziani, and Serena Favarin.
\newblock Exploring the {Effect} of 2019-{nCoV} {Containment} {Policies} on
  {Crime}: {The} {Case} of {Los} {Angeles}.
\newblock preprint, Open Science Framework, March 2020.
\newblock URL \url{https://osf.io/gcpq8}.

\bibitem[Mohler et~al.(2020)Mohler, Bertozzi, Carter, Short, Sledge, Tita,
  Uchida, and Brantingham]{mohler_impact_2020}
George Mohler, Andrea~L. Bertozzi, Jeremy Carter, Martin~B. Short, Daniel
  Sledge, George~E. Tita, Craig~D. Uchida, and P.~Jeffrey Brantingham.
\newblock Impact of social distancing during {COVID}-19 pandemic on crime in
  {Los} {Angeles} and {Indianapolis}.
\newblock \emph{Journal of Criminal Justice}, 68:\penalty0 101692, May 2020.
\newblock ISSN 0047-2352.
\newblock \doi{10.1016/j.jcrimjus.2020.101692}.
\newblock URL
  \url{http://www.sciencedirect.com/science/article/pii/S0047235220301860}.

\bibitem[Ashby(2020)]{ashby_initial_2020}
Matthew P.~J. Ashby.
\newblock Initial evidence on the relationship between the coronavirus pandemic
  and crime in the {United} {States}.
\newblock \emph{Crime Science}, 9\penalty0 (1):\penalty0 6, December 2020.
\newblock ISSN 2193-7680.
\newblock \doi{10.1186/s40163-020-00117-6}.
\newblock URL
  \url{https://crimesciencejournal.biomedcentral.com/articles/10.1186/s40163-020-00117-6}.

\bibitem[Abrams(2020)]{abrams_covid_2020}
David Abrams.
\newblock {COVID} and {Crime}: {An} {Early} {Empirical} {Look}.
\newblock {SSRN} {Scholarly} {Paper} ID 3674032, Social Science Research
  Network, Rochester, NY, August 2020.
\newblock URL \url{https://papers.ssrn.com/abstract=3674032}.

\bibitem[Rosenfeld et~al.(2021)Rosenfeld, Abt, and
  Lopez]{rosenfeld_pandemic_2020}
Richard Rosenfeld, Thomas Abt, and Ernesto Lopez.
\newblock Pandemic, {Social} {Unrest}, and {Crime} in {U}.{S}. {Cities}: {2020}
  {Year-End} {Update}.
\newblock \emph{National Commission on COVID-19 and Criminal Justice}, January
  2021.
\newblock URL
  \url{https://cdn.ymaws.com/counciloncj.org/resource/resmgr/covid_commission/Year_End_Crime_Update_Design.pdf}.

\bibitem[Coote(2020)]{coote_fbi_2020}
Darryl Coote.
\newblock {FBI}: {Murders} climbed amid a decrease in violent crime in 2020,
  September 2020.
\newblock URL
  \url{https://www.upi.com/Top_News/US/2020/09/15/FBI-Murders-climbed-amid-a-decrease-in-violent-crime-in-2020/9471600225926/}.

\bibitem[WorldPopulationReview.com(2020)]{worldpopulationreviewcom_200_2020}
WorldPopulationReview.com.
\newblock The 200 {Largest} {Cities} in the {United} {States} by {Population}
  2020, 2020.
\newblock URL \url{https://worldpopulationreview.com/us-cities}.

\bibitem[Bureau(2020)]{us_census_bureau_city_2020}
U.S.~Census Bureau.
\newblock City and {Town} {Population} {Totals}: 2010-2019: {Subcounty}
  {Resident} {Population} {Estimates}: {April} 1, 2010 to {July} 1, 2019
  ({SUB}-{EST2019}), 2020.
\newblock URL
  \url{https://www.census.gov/data/datasets/time-series/demo/popest/2010s-total-cities-and-towns.html}.
\newblock Section: Government.

\bibitem[CNN(2020)]{cnn_this_2020}
CNN.
\newblock This is where each state is during its phased reopening, 2020.
\newblock URL
  \url{https://www.cnn.com/interactive/2020/us/states-reopen-coronavirus-trnd/}.

\bibitem[Times(2020)]{the_new_york_times_see_2020}
The New~York Times.
\newblock See {Coronavirus} {Restrictions} and {Mask} {Mandates} for {All} 50
  {States}.
\newblock \emph{The New York Times}, April 2020.
\newblock ISSN 0362-4331.
\newblock URL
  \url{https://www.nytimes.com/interactive/2020/us/states-reopen-map-coronavirus.html}.

\end{thebibliography}

\section*{Appendix A}

\begin{table}[h]
\centering
\begin{tabular}{|c|c|c|c|}
\hline
City & \begin{tabular}[c]{@{}c@{}}Used in Seasonal Crime\\  Rates  Model\end{tabular} & \begin{tabular}[c]{@{}c@{}}Used in Weeks \\ into Lockdown Models\end{tabular} & \begin{tabular}[c]{@{}c@{}}Used in Openness\\  Index Model\end{tabular} \\ \hline
Atlanta & x & x & x \\ \hline
Austin & x & x & x \\ \hline
Baltimore & x & x & x \\ \hline
Boston & x & x & x \\ \hline
Chicago & x & x & x \\ \hline
Cincinnati & x & x & x \\ \hline
Dallas & x & x & x \\ \hline
Denver & x & x & x \\ \hline
Detroit & x & x & x \\ \hline
Houston & x & x & x \\ \hline
Kansas City* & x & x & x \\ \hline
Lincoln &  & x & x \\ \hline
Los Angeles & x & x & x \\ \hline
Louisville & x & x & x \\ \hline
Mesa & x & x &  \\ \hline
Milwaukee & x & x &  \\ \hline
Nashville & x & x & x \\ \hline
New York & x & x & x \\ \hline
Philadelphia & x & x & x \\ \hline
Phoenix & x & x & x \\ \hline
Pittsburgh & x & x & x \\ \hline
Raleigh & x & x &  \\ \hline
Sacramento & x & x & x \\ \hline
Saint Louis & x & x & x \\ \hline
Saint Paul & x & x &  \\ \hline
San Francisco** & x & x & x \\ \hline
Seattle & x & x & x \\ \hline
Tulsa & x & x &  \\ \hline
Washington & x & x & x \\ \hline
\end{tabular}
    \caption{Cities used in our analyses}
    \label{table:a2}
\end{table}

\textbf{Notes about Table 1:}
\begin{itemize}
    \item *Crime rates for weeks 0-11 of 2019 were unavailable for Kansas City
    \item **San Francisco not included in homicide models because weekly homicide counts not available
\end{itemize}

\begin{table}[h]
    \centering
    \begin{tabular}{|p{3cm}|p{2cm}|p{2cm}|p{2cm}|p{2cm}|p{2cm}|p{2cm}|}
    \hline
        \textbf{Outdoor and Recreation} & \textbf{Industries} & \textbf{Retail} & \textbf{Food and Drink} & \textbf{Entertainment} & \textbf{Personal Care} & \textbf{Places of Worship} \\ \hline
        State Parks & Construction & Retail Stores & Farmers Markets & Libraries & Pet Grooming & Church \\ \hline
        Fishing & Manufacturing & Malls & Restaurants & Movie Theaters & Hair Salons &  \\ \hline
        Boating & Agriculture & Apparel & Bars & Bowling alleys & Barbershops &  \\ \hline
        Golf Courses & Forestry & Electronics and appliances & Breweries & Museums & Nail Salons &  \\ \hline
        Marinas & Fishing & Florists & Clubs & Casinos & Personal Services &  \\ \hline
        Campgrounds & Hunting & Luggage &  & Zoos & Tanning Salons &  \\ \hline
        Beaches & Offices &  &  & Aquariums & Tattoo Parlors &  \\ \hline
        Gyms & Education &  &  & Entertainment Venues &  &  \\ \hline
        Pools &  &  &  &  &  &  \\ \hline
        Spas &  &  &  &  &  &  \\ \hline
    \end{tabular}
    \caption{Categories used for our openness index and which entities fell under which category}
    \label{tabel:a3}
\end{table}

\begin{table}[h]
    \centering
    \begin{tabular}{|l|l|}
    \hline
        Month & Number of Weeks \\ \hline
        1 & 4 \\ \hline
        2 & 4 \\ \hline
        3 & 4 \\ \hline
        4 & 5 \\ \hline
        5 & 4 \\ \hline
        6 & 4 \\ \hline
        7 & 5 \\ \hline
        8 & 4 \\ \hline
        9 & 5 \\ \hline
        10 & 4 \\ \hline
        11 & 4 \\ \hline
        12 & 4 \\ \hline
    \end{tabular}
    \caption{Number of weeks in each month grouping; same month groupings used for each year}
    \label{table:a4}
\end{table}

\begin{figure}[h]
    \centering
    \includegraphics{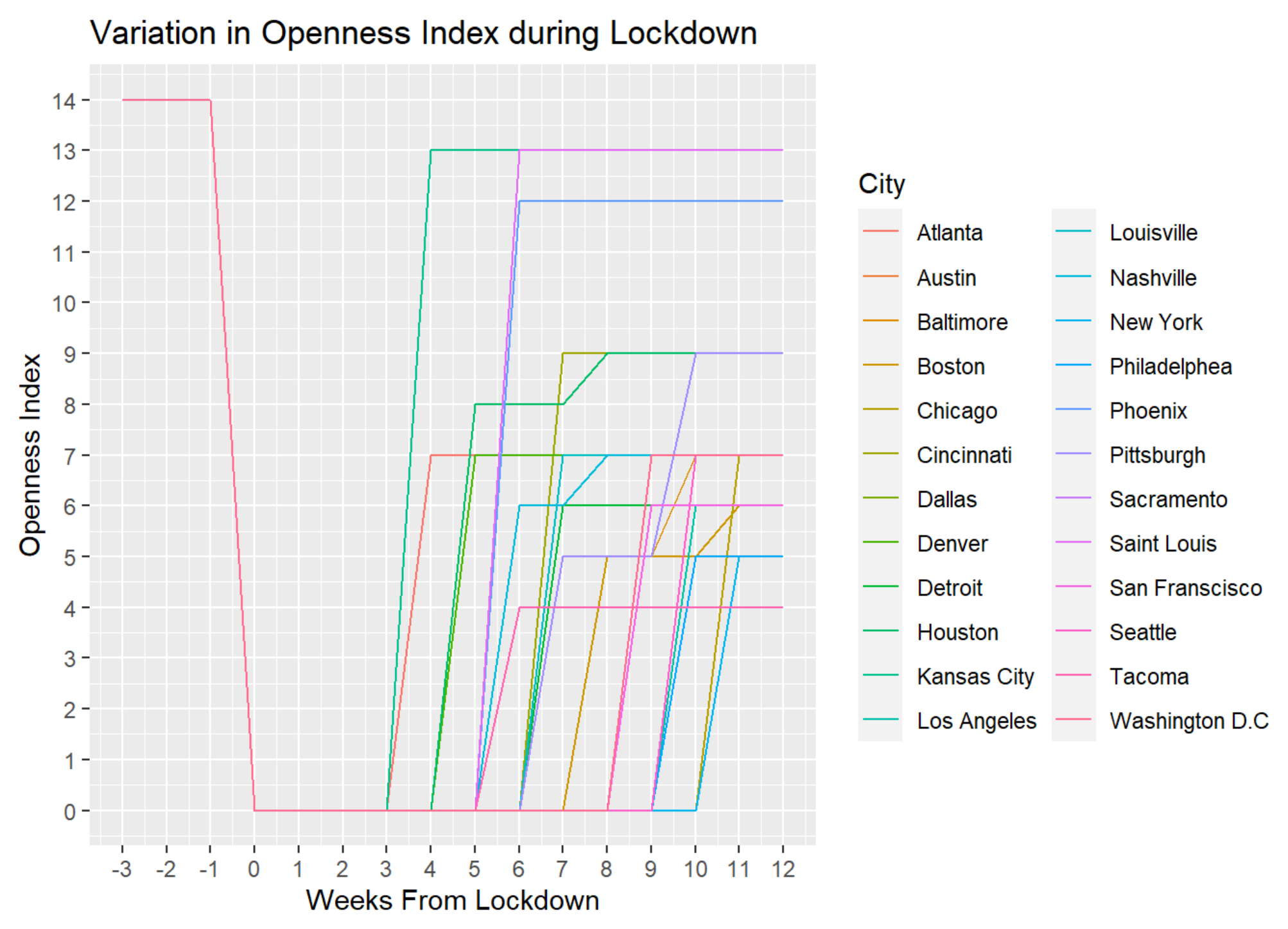}
    \caption{Variation in the openness index for each city}
    \label{fig:open}
\end{figure}

\clearpage

\section*{Appendix B}

Signif. codes throughout Appendix B:  ‘***’: 0.001;  ‘**’: 0.01; ‘*’: 0.05; ‘.’: 0.1 

\begin{table}[h]
\centering
\begin{tabular}{|l|l|l|l|ll|}
\hline
\textbf{} & Estimate & Std. Error & t value & P-value &  \\ \hline
(Intercept) & -1.147 & 0.705 & -1.626 & 0.104 & . \\ \hline
Year 2019 & 1.024 & 0.506 & 2.022 & 0.043 & * \\ \hline
Month 2 & -3.391 & 1.743 & -1.945 & 0.052 & . \\ \hline
Month 3 & -2.442 & 1.183 & -2.065 & 0.039 & * \\ \hline
Month 4 & 0.441 & 0.859 & 0.514 & 0.608 &  \\ \hline
Month 5 & 3.551 & 1.228 & 2.891 & 0.004 & ** \\ \hline
Month 6 & 3.871 & 1.313 & 2.947 & 0.003 & ** \\ \hline
Month 7 & 5.564 & 1.545 & 3.601 & 0.000 & *** \\ \hline
Month 8 & 3.228 & 1.339 & 2.410 & 0.016 & * \\ \hline
Month 9 & 2.018 & 1.305 & 1.546 & 0.122 &  \\ \hline
Month 10 & 1.300 & 1.431 & 0.909 & 0.363 &  \\ \hline
Month 11 & -0.155 & 1.392 & -0.111 & 0.911 &  \\ \hline
Month 12 & -1.370 & 1.445 & -0.948 & 0.343 &  \\ \hline
PrePandemic & 4.085 & 3.006 & 1.359 & 0.174 &  \\ \hline
Lockdown & 3.334 & 1.547 & 2.156 & 0.031 & * \\ \hline
Summer Protests & 11.246 & 3.301 & 3.407 & 0.001 & *** \\ \hline
End of Year & 7.292 & 1.374 & 5.306 & 0.000 & *** \\ \hline
\end{tabular}
\caption{Homicide Seasonal Effects Model Results}
    \label{table:hom-month}
\end{table}

\begin{figure}[h]
    \centering
    \includegraphics[scale=0.75]{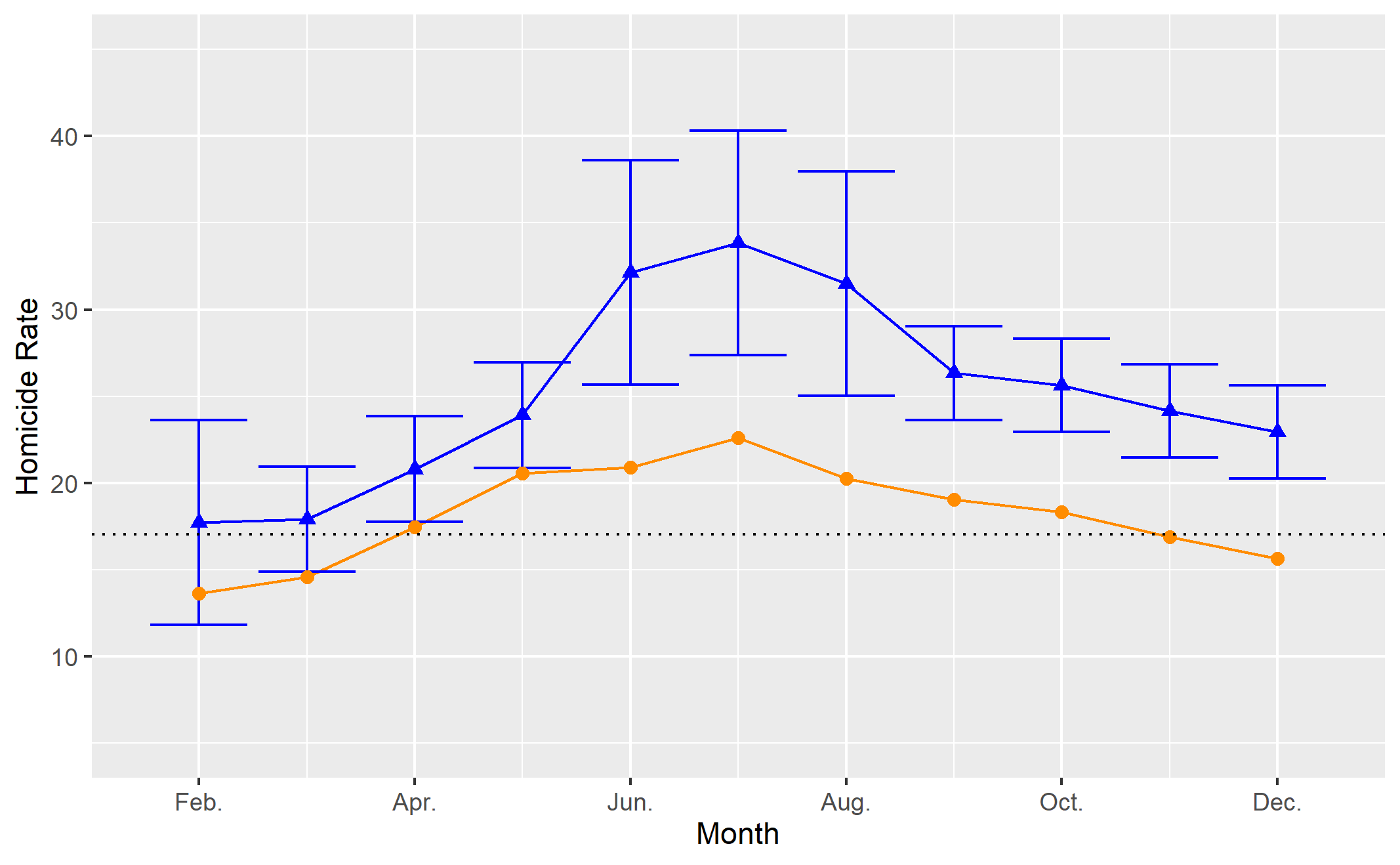}
    \caption{Homicide seasonal trends in 2018 and 2019 (orange) and in 2020 (blue) based on seasonal model results with 95\% confidence intervals}
    \label{fig:hom_month_oldfig1}
\end{figure}

\begin{table}[h]
\centering
\begin{tabular}{|l|l|l|l|l|}
\hline
Effect & Difference From PrePandemic Effect & Std. Error & t-stat & P-value \\ \hline
Lockdown & -0.751 & 2.135 & -0.352 & 0.725 \\ \hline
Summer Protests & 7.161 & 3.698 & 1.936 & 0.053 \\ \hline
End of Year & 3.207 & 2.719 & 1.180 & 0.238 \\ \hline
\end{tabular}
\caption{Homicide Post-Estimation Test Results}
\label{table:hom-postest}
\end{table}

\begin{table}[h]
    \centering
    \begin{tabular}{|l|l|l|l|ll|}
    \hline
         & Estimate & Std. Error & t value & P-value &  \\ \hline
        (Intercept) & -0.913 & 0.740 & -1.234 & 0.217 &  \\ \hline
        Year 2019 & 0.885 & 0.489 & 1.809 & 0.070 & . \\ \hline
        Year 2020 & 3.130 & 1.942 & 1.612 & 0.107 &  \\ \hline
        Month 2 & -3.242 & 1.648 & -1.968 & 0.049 & * \\ \hline
        Month 3 & -2.167 & 1.160 & -1.868 & 0.062 & . \\ \hline
        Month 4 & -0.215 & 1.157 & -0.186 & 0.853 &  \\ \hline
        Month 5 & 3.325 & 1.112 & 2.990 & 0.003 & ** \\ \hline
        Month 6 & 2.850 & 1.045 & 2.727 & 0.006 & ** \\ \hline
        Month 7 & 4.367 & 1.136 & 3.844 & 0.000 & *** \\ \hline
        Month 8 & 4.253 & 1.413 & 3.010 & 0.003 & ** \\ \hline
        Month 9 & 1.629 & 1.434 & 1.136 & 0.256 &  \\ \hline
        Month 10 & 1.321 & 1.153 & 1.146 & 0.252 &  \\ \hline
        Month 11 & -0.916 & 1.818 & -0.504 & 0.614 &  \\ \hline
        Month 12 & -1.005 & 1.290 & -0.779 & 0.436 &  \\ \hline
        Week Before \& Week of Lockdown & -1.566 & 1.201 & -1.304 & 0.192 &  \\ \hline
        One to Two weeks after Lockdown Start & -1.368 & 3.268 & -0.419 & 0.675 &  \\ \hline
        Three to Four weeks after Lockdown Start & 4.205 & 3.835 & 1.096 & 0.273 &  \\ \hline
        Five to Six weeks after Lockdown Start & 0.724 & 1.759 & 0.412 & 0.681 &  \\ \hline
        Seven to Eight weeks after Lockdown Start & -2.489 & 2.206 & -1.129 & 0.259 &  \\ \hline
        Nine to Ten weeks after Lockdown Start & 2.733 & 3.786 & 0.722 & 0.470 &  \\ \hline
    \end{tabular}
    \caption{Homicide Biweekly Weeks since Lockdown Model Results}
    \label{table:hom-biweek}
\end{table}

\begin{table}[h]
    \centering
    \begin{tabular}{|l|l|l|l|ll|}
    \hline
         & Estimate & Std. Error & t value & P-value &  \\ \hline
        (Intercept) & -4.124 & 3.360 & -1.227 & 0.220 &  \\ \hline
        Year 2019 & 0.989 & 0.562 & 1.761 & 0.078 & . \\ \hline
        Year 2020 & 4.762 & 3.344 & 1.424 & 0.154 &  \\ \hline
        Month 2 & -4.312 & 1.927 & -2.238 & 0.025 & * \\ \hline
        Month 3 & -2.946 & 1.150 & -2.561 & 0.010 & * \\ \hline
        Month 4 & 0.473 & 0.633 & 0.746 & 0.456 &  \\ \hline
        Month 5 & 3.684 & 1.331 & 2.768 & 0.006 & ** \\ \hline
        Month 6 & 3.307 & 1.189 & 2.782 & 0.005 & ** \\ \hline
        Month 7 & 5.082 & 1.411 & 3.602 & 0.000 & *** \\ \hline
        Month 8 & 4.434 & 1.850 & 2.396 & 0.017 & * \\ \hline
        Month 9 & 1.865 & 1.554 & 1.200 & 0.230 &  \\ \hline
        Month 10 & 1.131 & 1.158 & 0.976 & 0.329 &  \\ \hline
        Month 11 & -1.614 & 1.769 & -0.912 & 0.362 &  \\ \hline
        Month 12 & -1.202 & 1.165 & -1.032 & 0.302 &  \\ \hline
        Openness Index & 0.229 & 0.245 & 0.932 & 0.351 &  \\ \hline
    \end{tabular}
    \caption{Homicide Openness Index Model Results}
    \label{table:hom-open}
\end{table}

\begin{table}[h]
\centering
\begin{tabular}{|l|l|l|l|ll|}
\hline
\textbf{} & Estimate & Std. Error & t value & P-value &  \\ \hline
(Intercept) & 7.035 & 19.071 & 0.369 & 0.712 &  \\ \hline
Year 2019 & -24.446 & 21.953 & -1.114 & 0.265 &  \\ \hline
Month 2 & -30.787 & 8.181 & -3.763 & 0.000 & *** \\ \hline
Month 3 & -58.931 & 18.049 & -3.265 & 0.001 & ** \\ \hline
Month 4 & -45.174 & 23.213 & -1.946 & 0.052 & . \\ \hline
Month 5 & -11.305 & 15.127 & -0.747 & 0.455 &  \\ \hline
Month 6 & -5.110 & 26.780 & -0.191 & 0.849 &  \\ \hline
Month 7 & 23.726 & 25.285 & 0.938 & 0.348 &  \\ \hline
Month 8 & 21.159 & 23.875 & 0.886 & 0.375 &  \\ \hline
Month 9 & 5.830 & 24.664 & 0.236 & 0.813 &  \\ \hline
Month 10 & 4.518 & 23.568 & 0.192 & 0.848 &  \\ \hline
Month 11 & 4.028 & 25.923 & 0.155 & 0.877 &  \\ \hline
Month 12 & 6.253 & 28.534 & 0.219 & 0.827 &  \\ \hline
PrePandemic & -16.329 & 34.273 & -0.476 & 0.634 &  \\ \hline
Lockdown & -6.939 & 34.119 & -0.203 & 0.839 &  \\ \hline
Summer Protests & 85.592 & 32.269 & 2.652 & 0.008 & ** \\ \hline
End of Year & 114.826 & 38.889 & 2.953 & 0.003 & ** \\ \hline
\end{tabular}
    \caption{Auto Theft Seasonal Effects Model Results}
    \label{table:auto-month}
\end{table}

\begin{table}[h]
\centering
\begin{tabular}{|l|l|l|l|l|}
\hline
Effect & Difference From PrePandemic Effect & Std. Error & t-stat & P-value \\ \hline
Lockdown & 9.390 & 19.472 & 0.482 & 0.630 \\ \hline
Summer Protests & 101.921 & 36.551 & 2.788 & 0.005 \\ \hline
End of Year & 131.155 & 43.188 & 3.037 & 0.002 \\ \hline
\end{tabular}
\caption{Auto Theft Post-Estimation Test Results}
\label{table:auto-postest}
\end{table}

\begin{table}[h]
    \centering
    \begin{tabular}{|l|l|l|l|ll|}
    \hline
         & Estimate & Std. Error & t value & P-value &  \\ \hline
        (Intercept) & 11.379 & 19.229 & 0.592 & 0.554 &  \\ \hline
        Year 2019 & -23.917 & 21.142 & -1.131 & 0.258 &  \\ \hline
        Year 2020 & -30.261 & 38.200 & -0.792 & 0.428 &  \\ \hline
        Month 2 & -30.283 & 7.950 & -3.809 & 0.000 & *** \\ \hline
        Month 3 & -58.228 & 16.071 & -3.623 & 0.000 & *** \\ \hline
        Month 4 & -54.254 & 24.476 & -2.217 & 0.027 & * \\ \hline
        Month 5 & -22.654 & 18.887 & -1.199 & 0.230 &  \\ \hline
        Month 6 & -2.346 & 26.473 & -0.089 & 0.929 &  \\ \hline
        Month 7 & 17.530 & 25.048 & 0.700 & 0.484 &  \\ \hline
        Month 8 & 11.055 & 23.671 & 0.467 & 0.640 &  \\ \hline
        Month 9 & 9.153 & 24.847 & 0.368 & 0.713 &  \\ \hline
        Month 10 & -2.170 & 23.839 & -0.091 & 0.927 &  \\ \hline
        Month 11 & -1.781 & 26.433 & -0.067 & 0.946 &  \\ \hline
        Month 12 & -4.207 & 27.653 & -0.152 & 0.879 &  \\ \hline
        Week Before \& Week of Lockdown & 28.601 & 26.816 & 1.067 & 0.286 &  \\ \hline
        One to Two weeks after Lockdown Start & 32.120 & 28.535 & 1.126 & 0.260 &  \\ \hline
        Three to Four weeks after Lockdown Start & 69.694 & 36.919 & 1.888 & 0.059 & . \\ \hline
        Five to Six weeks after Lockdown Start & 27.905 & 34.964 & 0.798 & 0.425 &  \\ \hline
        Seven to Eight weeks after Lockdown Start & 24.336 & 37.187 & 0.654 & 0.513 &  \\ \hline
        Nine to Ten weeks after Lockdown Start & 106.100 & 46.256 & 2.294 & 0.022 & * \\ \hline
    \end{tabular}
    \caption{Auto Theft Biweekly Weeks since Lockdown Model Results}
    \label{table:auto-biweek}
\end{table}

\begin{table}[h]
    \centering
    \begin{tabular}{|l|l|l|l|ll|}
    \hline
         & Estimate & Std. Error & t value & P-value &  \\ \hline
        (Intercept) & 31.030 & 56.145 & 0.553 & 0.580 &  \\ \hline
        Year 2019 & -21.266 & 24.182 & -0.879 & 0.379 &  \\ \hline
        Year 2020 & -24.608 & 48.506 & -0.507 & 0.612 &  \\ \hline
        Month 2 & -26.476 & 7.523 & -3.519 & 0.000 & *** \\ \hline
        Month 3 & -48.278 & 13.749 & -3.511 & 0.000 & *** \\ \hline
        Month 4 & -42.203 & 30.134 & -1.400 & 0.161 &  \\ \hline
        Month 5 & -15.264 & 18.455 & -0.827 & 0.408 &  \\ \hline
        Month 6 & 5.092 & 28.103 & 0.181 & 0.856 &  \\ \hline
        Month 7 & 32.003 & 25.478 & 1.256 & 0.209 &  \\ \hline
        Month 8 & 20.277 & 26.575 & 0.763 & 0.445 &  \\ \hline
        Month 9 & 27.070 & 24.179 & 1.120 & 0.263 &  \\ \hline
        Month 10 & 4.426 & 24.977 & 0.177 & 0.859 &  \\ \hline
        Month 11 & 2.399 & 31.984 & 0.075 & 0.940 &  \\ \hline
        Month 12 & -2.371 & 34.895 & -0.068 & 0.946 &  \\ \hline
        Openness Index & -1.997 & 2.723 & -0.733 & 0.463 &  \\ \hline
    \end{tabular}
    \caption{Auto Theft Openness Index Model Results}
    \label{table:auto-open}
\end{table}

\begin{table}[h]
\centering
\begin{tabular}{|l|l|l|l|ll|}
\hline
\textbf{} & Estimate & Std. Error & t value & P-value &  \\ \hline
(Intercept) & -13.573 & 13.832 & -0.981 & 0.326 &  \\ \hline
Year 2019 & -53.910 & 22.544 & -2.391 & 0.017 & * \\ \hline
Month 2 & -38.055 & 8.370 & -4.547 & 0.000 & *** \\ \hline
Month 3 & -60.853 & 12.409 & -4.904 & 0.000 & *** \\ \hline
Month 4 & -14.586 & 9.515 & -1.533 & 0.125 &  \\ \hline
Month 5 & 33.267 & 20.023 & 1.661 & 0.097 & . \\ \hline
Month 6 & 70.519 & 26.076 & 2.704 & 0.007 & ** \\ \hline
Month 7 & 51.925 & 22.924 & 2.265 & 0.024 & * \\ \hline
Month 8 & 57.516 & 21.971 & 2.618 & 0.009 & ** \\ \hline
Month 9 & 27.064 & 20.082 & 1.348 & 0.178 &  \\ \hline
Month 10 & 27.435 & 19.267 & 1.424 & 0.154 &  \\ \hline
Month 11 & 0.190 & 24.574 & 0.008 & 0.994 &  \\ \hline
Month 12 & 2.538 & 24.666 & 0.103 & 0.918 &  \\ \hline
PrePandemic & -87.751 & 35.289 & -2.487 & 0.013 & * \\ \hline
Lockdown & -69.521 & 42.859 & -1.622 & 0.105 &  \\ \hline
Summer Protests & -74.699 & 42.972 & -1.738 & 0.082 & . \\ \hline
End of Year & -71.007 & 42.882 & -1.656 & 0.098 & . \\ \hline
\end{tabular}
    \caption{Burglary Seasonal Effects Model Results}
    \label{table:burg-month}
\end{table}

\begin{table}[h]
\centering
\begin{tabular}{|l|l|l|l|l|}
\hline
Effect & Difference From PrePandemic Effect & Std. Error & t-stat & P-value \\ \hline
Lockdown & 18.229 & 28.285 & 0.644 & 0.519 \\ \hline
Summer Protests & 13.051 & 30.138 & 0.433 & 0.665 \\ \hline
End of Year & 16.743 & 36.838 & 0.455 & 0.649 \\ \hline
\end{tabular}
\caption{Burglary Post-Estimation Test Results}
\label{table:burg-postest}
\end{table}

\begin{table}[h]
    \centering
    \begin{tabular}{|l|l|l|l|ll|}
    \hline
         & Estimate & Std. Error & t value & P-value &  \\ \hline
        (Intercept) & -11.027 & 14.029 & -0.786 & 0.432 &  \\ \hline
        Year 2019 & -54.313 & 21.757 & -2.496 & 0.013 & * \\ \hline
        Year 2020 & -99.531 & 37.057 & -2.686 & 0.007 & ** \\ \hline
        Month 2 & -36.957 & 8.066 & -4.582 & 0.000 & *** \\ \hline
        Month 3 & -56.510 & 10.157 & -5.564 & 0.000 & *** \\ \hline
        Month 4 & -11.792 & 10.404 & -1.133 & 0.257 &  \\ \hline
        Month 5 & 15.368 & 19.147 & 0.803 & 0.422 &  \\ \hline
        Month 6 & 27.617 & 21.273 & 1.298 & 0.194 &  \\ \hline
        Month 7 & 68.507 & 23.262 & 2.945 & 0.003 & ** \\ \hline
        Month 8 & 68.634 & 21.212 & 3.236 & 0.001 & ** \\ \hline
        Month 9 & 27.671 & 19.813 & 1.397 & 0.163 &  \\ \hline
        Month 10 & 19.427 & 19.212 & 1.011 & 0.312 &  \\ \hline
        Month 11 & -2.670 & 25.998 & -0.103 & 0.918 &  \\ \hline
        Month 12 & 0.206 & 23.758 & 0.009 & 0.993 &  \\ \hline
        Week Before \& Week of Lockdown & 41.128 & 24.847 & 1.655 & 0.098 & . \\ \hline
        One to Two weeks after Lockdown Start & 31.519 & 28.444 & 1.108 & 0.268 &  \\ \hline
        Three to Four weeks after Lockdown Start & 18.600 & 36.358 & 0.512 & 0.609 &  \\ \hline
        Five to Six weeks after Lockdown Start & -26.413 & 35.751 & -0.739 & 0.460 &  \\ \hline
        Seven to Eight weeks after Lockdown Start & -9.734 & 46.880 & -0.208 & 0.836 &  \\ \hline
        Nine to Ten weeks after Lockdown Start & 405.358 & 230.724 & 1.757 & 0.079 &  \\ \hline
    \end{tabular}
    \caption{Burglary Biweekly Weeks since Lockdown Model Results}
    \label{table:burg-biweek}
\end{table}

\begin{figure}[h]
    \centering
    \includegraphics[scale=0.75]{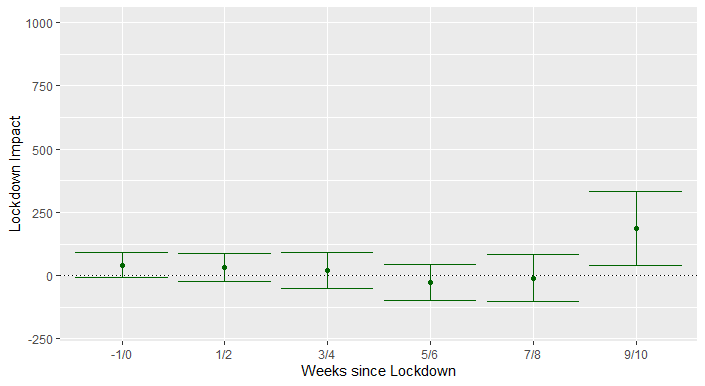}
    \caption{Biweekly 2020 differences in burglary rates during the lockdown relative to typical seasonal levels with 95\% confidence intervals; St. Paul, week 21 removed to analyze the sensitivity of results to this outlier}
    \label{fig:burg_no_stp}
\end{figure}

\begin{table}[h]
    \centering
    \begin{tabular}{|l|l|l|l|ll|}
    \hline
         & Estimate & Std. Error & t value & P-value &  \\ \hline
        (Intercept) & -11.027 & 14.029 & -0.786 & 0.432 &  \\ \hline
        Year 2019 & -54.313 & 21.757 & -2.496 & 0.013 & * \\ \hline
        Year 2020 & -99.531 & 37.057 & -2.686 & 0.007 & ** \\ \hline
        Month 2 & 19.427 & 19.212 & 1.011 & 0.312 &  \\ \hline
        Month 3 & -2.670 & 25.998 & -0.103 & 0.918 &  \\ \hline
        Month 4 & 0.206 & 23.758 & 0.009 & 0.993 &  \\ \hline
        Month 5 & -36.957 & 8.066 & -4.582 & 0.000 & *** \\ \hline
        Month 6 & -56.510 & 10.157 & -5.564 & 0.000 & *** \\ \hline
        Month 7 & -11.792 & 10.404 & -1.133 & 0.257 &  \\ \hline
        Month 8 & 15.368 & 19.147 & 0.803 & 0.422 &  \\ \hline
        Month 9 & 27.617 & 21.273 & 1.298 & 0.194 &  \\ \hline
        Month 10 & 68.507 & 23.262 & 2.945 & 0.003 & ** \\ \hline
        Month 11 & 68.634 & 21.212 & 3.236 & 0.001 & ** \\ \hline
        Month 12 & 27.671 & 19.813 & 1.397 & 0.163 &  \\ \hline
        Week Before \& Week of Lockdown & 41.128 & 24.847 & 1.655 & 0.098 & . \\ \hline
        One to Two weeks after Lockdown Start & 31.519 & 28.444 & 1.108 & 0.268 &  \\ \hline
        Three to Four weeks after Lockdown Start & 18.600 & 36.358 & 0.512 & 0.609 &  \\ \hline
        Five to Six weeks after Lockdown Start & -26.413 & 35.751 & -0.739 & 0.460 &  \\ \hline
        Seven to Eight weeks after Lockdown Start & -9.734 & 46.880 & -0.208 & 0.836 &  \\ \hline
        Nine to Ten weeks after Lockdown Start & 185.361 & 74.723 & 2.481 & 0.013 & * \\ \hline
    \end{tabular}
    \caption{Burglary Biweekly Weeks since Lockdown Model Results; St. Paul, week 21 removed}
    \label{table:burg-biweek-nostp}
\end{table}

\begin{table}[h]
    \centering
    \begin{tabular}{|l|l|l|l|ll|}
    \hline
         & Estimate & Std. Error & t value & P-value &  \\ \hline
        (Intercept) & 66.895 & 66.602 & 1.004 & 0.315 &  \\ \hline
        Year 2019 & -58.035 & 25.275 & -2.296 & 0.022 & * \\ \hline
        Year 2020 & -116.029 & 51.601 & -2.249 & 0.025 & * \\ \hline
        Month 2 & -32.052 & 8.469 & -3.785 & 0.000 & *** \\ \hline
        Month 3 & -52.185 & 11.044 & -4.725 & 0.000 & *** \\ \hline
        Month 4 & -28.551 & 18.436 & -1.549 & 0.121 &  \\ \hline
        Month 5 & 10.104 & 22.156 & 0.456 & 0.648 &  \\ \hline
        Month 6 & 16.925 & 26.433 & 0.640 & 0.522 &  \\ \hline
        Month 7 & 52.912 & 26.291 & 2.013 & 0.044 & * \\ \hline
        Month 8 & 50.397 & 21.779 & 2.314 & 0.021 & * \\ \hline
        Month 9 & 12.400 & 22.736 & 0.545 & 0.585 &  \\ \hline
        Month 10 & 17.090 & 22.813 & 0.749 & 0.454 &  \\ \hline
        Month 11 & -9.185 & 33.005 & -0.278 & 0.781 &  \\ \hline
        Month 12 & -10.179 & 30.583 & -0.333 & 0.739 &  \\ \hline
        Openness Index & -4.985 & 3.741 & -1.332 & 0.183 &  \\ \hline
    \end{tabular}
    \caption{Burglary Openness Index Model Results}
    \label{table:burg-open}
\end{table}

\begin{table}[h]
\centering
\begin{tabular}{|l|l|l|l|ll|}
\hline
\textbf{} & Estimate & Std. Error & t value & P-value &  \\ \hline
(Intercept) & -1.388 & 5.853 & -0.237 & 0.813 &  \\ \hline
Year 2019 & -17.090 & 10.105 & -1.691 & 0.091 & . \\ \hline
Month 2 & -29.167 & 6.126 & -4.761 & 0.000 & *** \\ \hline
Month 3 & -31.902 & 7.489 & -4.260 & 0.000 & *** \\ \hline
Month 4 & -33.838 & 9.211 & -3.674 & 0.000 & *** \\ \hline
Month 5 & -3.802 & 4.290 & -0.886 & 0.376 &  \\ \hline
Month 6 & 12.122 & 7.463 & 1.624 & 0.104 &  \\ \hline
Month 7 & 28.766 & 8.187 & 3.514 & 0.000 & *** \\ \hline
Month 8 & 31.908 & 9.248 & 3.450 & 0.001 & *** \\ \hline
Month 9 & 23.258 & 8.119 & 2.865 & 0.004 & ** \\ \hline
Month 10 & 15.869 & 9.851 & 1.611 & 0.107 &  \\ \hline
Month 11 & 6.574 & 12.386 & 0.531 & 0.596 &  \\ \hline
Month 12 & -6.640 & 14.621 & -0.454 & 0.650 &  \\ \hline
PrePandemic & -24.727 & 17.059 & -1.450 & 0.147 &  \\ \hline
Lockdown & -54.763 & 14.939 & -3.666 & 0.000 & *** \\ \hline
Summer Protests & -61.961 & 20.241 & -3.061 & 0.002 & ** \\ \hline
End of Year & -48.646 & 18.811 & -2.586 & 0.010 & ** \\ \hline
\end{tabular}
    \caption{Robbery Seasonal Effects Model Results}
    \label{table:rob-month}
\end{table}

\begin{table}[]
\centering
\begin{tabular}{|l|l|l|l|l|}
\hline
Effect & Difference From PrePandemic Effect & Std. Error & t-stat & P-value \\ \hline
Lockdown & -30.0355 & 10.9556 & -2.7416 & 0.0061 \\ \hline
Summer Protests & -37.2341 & 18.5049 & -2.0121 & 0.0443 \\ \hline
End of Year & -23.9193 & 17.4159 & -1.3734 & 0.1697 \\ \hline
\end{tabular}
\caption{Robbery Post-Estimation Test Results}
\label{table:rob-postest}
\end{table}

\begin{table}[h]
    \centering
    \begin{tabular}{|l|l|l|l|ll|}
    \hline
         & Estimate & Std. Error & t value & P-value &  \\ \hline
        (Intercept) & 0.358 & 5.864 & 0.061 & 0.951 &  \\ \hline
        Year 2019 & -16.520 & 9.770 & -1.691 & 0.091 & . \\ \hline
        Year 2020 & -29.933 & 18.337 & -1.632 & 0.103 &  \\ \hline
        Month 2 & -28.359 & 5.879 & -4.823 & 0.000 & *** \\ \hline
        Month 3 & -41.950 & 7.769 & -5.400 & 0.000 & *** \\ \hline
        Month 4 & -32.378 & 9.774 & -3.313 & 0.001 & *** \\ \hline
        Month 5 & -2.928 & 5.714 & -0.512 & 0.608 &  \\ \hline
        Month 6 & 4.942 & 8.065 & 0.613 & 0.540 &  \\ \hline
        Month 7 & 28.275 & 7.619 & 3.711 & 0.000 & *** \\ \hline
        Month 8 & 32.386 & 9.385 & 3.451 & 0.001 & *** \\ \hline
        Month 9 & 24.307 & 8.188 & 2.969 & 0.003 & ** \\ \hline
        Month 10 & 11.795 & 9.536 & 1.237 & 0.216 &  \\ \hline
        Month 11 & 0.428 & 12.680 & 0.034 & 0.973 &  \\ \hline
        Month 12 & -6.136 & 14.645 & -0.419 & 0.675 &  \\ \hline
        Week Before \& Week of Lockdown & 17.939 & 14.008 & 1.281 & 0.200 &  \\ \hline
        One to Two weeks after Lockdown Start & -26.977 & 17.546 & -1.538 & 0.124 &  \\ \hline
        Three to Four weeks after Lockdown Start & -37.246 & 17.083 & -2.180 & 0.029 & * \\ \hline
        Five to Six weeks after Lockdown Start & -45.540 & 14.231 & -3.200 & 0.001 & ** \\ \hline
        Seven to Eight weeks after Lockdown Start & -38.838 & 22.803 & -1.703 & 0.089 & . \\ \hline
        Nine to Ten weeks after Lockdown Start & -16.433 & 21.859 & -0.752 & 0.452 &  \\ \hline
    \end{tabular}
    \caption{Robbery Biweekly Weeks since Lockdown Model Results}
    \label{table:rob-biweek}
\end{table}

\begin{table}[h]
    \centering
    \begin{tabular}{|l|l|l|l|ll|}
    \hline
         & Estimate & Std. Error & t value & P-value &  \\ \hline
        (Intercept) & -33.941 & 27.844 & -1.219 & 0.223 &  \\ \hline
        Year 2019 & -15.218 & 11.645 & -1.307 & 0.191 &  \\ \hline
        Year 2020 & -31.880 & 22.883 & -1.393 & 0.164 &  \\ \hline
        Month 2 & -31.979 & 6.500 & -4.920 & 0.000 & *** \\ \hline
        Month 3 & -43.465 & 8.523 & -5.099 & 0.000 & *** \\ \hline
        Month 4 & -37.154 & 13.440 & -2.764 & 0.006 & ** \\ \hline
        Month 5 & -9.909 & 5.018 & -1.975 & 0.048 & * \\ \hline
        Month 6 & 2.879 & 9.967 & 0.289 & 0.773 &  \\ \hline
        Month 7 & 26.323 & 9.353 & 2.814 & 0.005 & ** \\ \hline
        Month 8 & 27.600 & 10.563 & 2.613 & 0.009 & ** \\ \hline
        Month 9 & 24.049 & 9.961 & 2.414 & 0.016 & * \\ \hline
        Month 10 & 9.586 & 11.980 & 0.800 & 0.424 &  \\ \hline
        Month 11 & -1.166 & 15.752 & -0.074 & 0.941 &  \\ \hline
        Month 12 & -11.419 & 18.371 & -0.622 & 0.534 &  \\ \hline
        Openness Index & 2.656 & 1.545 & 1.719 & 0.086 & . \\ \hline
    \end{tabular}
    \caption{Robbery Openness Index Model Results}
    \label{table:rob-open}
\end{table}

\begin{table}[]
\centering
\begin{tabular}{|l|l|l|l|ll|}
\hline
\textbf{} & Estimate & Std. Error & t value & P-value &  \\ \hline
(Intercept) & -172.839 & 42.468 & -4.070 & 0.000 & *** \\ \hline
Year 2019 & -75.770 & 81.451 & -0.930 & 0.352 &  \\ \hline
Month 2 & -80.650 & 24.359 & -3.311 & 0.001 & *** \\ \hline
Month 3 & -11.520 & 31.414 & -0.367 & 0.714 &  \\ \hline
Month 4 & 8.339 & 45.041 & 0.185 & 0.853 &  \\ \hline
Month 5 & 122.559 & 49.874 & 2.457 & 0.014 & * \\ \hline
Month 6 & 252.376 & 65.051 & 3.880 & 0.000 & *** \\ \hline
Month 7 & 378.880 & 69.899 & 5.420 & 0.000 & *** \\ \hline
Month 8 & 401.649 & 72.474 & 5.542 & 0.000 & *** \\ \hline
Month 9 & 297.203 & 60.137 & 4.942 & 0.000 & *** \\ \hline
Month 10 & 279.252 & 52.646 & 5.304 & 0.000 & *** \\ \hline
Month 11 & 181.866 & 51.404 & 3.538 & 0.000 & *** \\ \hline
Month 12 & 202.640 & 69.165 & 2.930 & 0.003 & ** \\ \hline
PrePandemic & 107.506 & 138.157 & 0.778 & 0.436 &  \\ \hline
Lockdown & -501.964 & 127.496 & -3.937 & 0.000 & *** \\ \hline
Summer Protests & -600.525 & 129.817 & -4.626 & 0.000 & *** \\ \hline
End of Year & -320.511 & 127.260 & -2.519 & 0.012 & * \\ \hline
\end{tabular}
    \caption{Larceny Seasonal Effects Model Results}
    \label{table:larceny-month}
\end{table}

\begin{table}[h]
\centering
\begin{tabular}{|l|l|l|l|l|}
\hline
Effect & Difference From PrePandemic Effect & Std. Error & t-stat & P-value \\ \hline
Lockdown & -609.469 & 87.024 & -7.003 & 0.000 \\ \hline
Summer Protests & -708.031 & 122.929 & -5.760 & 0.000 \\ \hline
End of Year & -428.017 & 117.077 & -3.656 & 0.000 \\ \hline
\end{tabular}
\caption{Larceny Post-Estimation Test Results}
\label{table:larceny-postest}
\end{table}

\begin{table}[h]
    \centering
    \begin{tabular}{|l|l|l|l|ll|}
    \hline
         & Estimate & Std. Error & t value & P-value &  \\ \hline
        (Intercept) & -165.197 & 47.533 & -3.475 & 0.001 & *** \\ \hline
        Year 2019 & -75.300 & 78.780 & -0.956 & 0.339 &  \\ \hline
        Year 2020 & -16.856 & 128.522 & -0.131 & 0.896 &  \\ \hline
        Month 2 & -72.312 & 23.655 & -3.057 & 0.002 & ** \\ \hline
        Month 3 & -126.390 & 40.801 & -3.098 & 0.002 & ** \\ \hline
        Month 4 & 29.778 & 53.010 & 0.562 & 0.574 &  \\ \hline
        Month 5 & 155.007 & 68.906 & 2.250 & 0.024 & * \\ \hline
        Month 6 & 267.398 & 69.673 & 3.838 & 0.000 & *** \\ \hline
        Month 7 & 396.858 & 74.144 & 5.353 & 0.000 & *** \\ \hline
        Month 8 & 357.109 & 76.742 & 4.653 & 0.000 & *** \\ \hline
        Month 9 & 337.438 & 61.448 & 5.491 & 0.000 & *** \\ \hline
        Month 10 & 285.273 & 55.251 & 5.163 & 0.000 & *** \\ \hline
        Month 11 & 118.234 & 56.930 & 2.077 & 0.038 & * \\ \hline
        Month 12 & 166.853 & 77.260 & 2.160 & 0.031 & * \\ \hline
        Week Before \& Week of Lockdown & -439.904 & 69.893 & -6.294 & 0.000 & *** \\ \hline
        One to Two weeks after Lockdown Start & -641.521 & 109.325 & -5.868 & 0.000 & *** \\ \hline
        Three to Four weeks after Lockdown Start & -590.909 & 114.231 & -5.173 & 0.000 & *** \\ \hline
        Five to Six weeks after Lockdown Start & -640.008 & 115.499 & -5.541 & 0.000 & *** \\ \hline
        Seven to Eight weeks after Lockdown Start & -678.235 & 138.096 & -4.911 & 0.000 & *** \\ \hline
        Nine to Ten weeks after Lockdown Start & -586.680 & 217.762 & -2.694 & 0.007 & ** \\ \hline
    \end{tabular}
    \caption{Larceny Biweekly Weeks since Lockdown Model Results}
    \label{table:larceny-biweek}
\end{table}

\begin{table}[h]
    \centering
    \begin{tabular}{|l|l|l|l|ll|}
    \hline
         & Estimate & Std. Error & t value & P-value &  \\ \hline
        (Intercept) & -716.418 & 237.643 & -3.015 & 0.003 & ** \\ \hline
        Year 2019 & -89.551 & 94.801 & -0.945 & 0.345 &  \\ \hline
        Year 2020 & -111.728 & 183.258 & -0.610 & 0.542 &  \\ \hline
        Month 2 & -76.206 & 27.483 & -2.773 & 0.006 & ** \\ \hline
        Month 3 & -162.573 & 54.426 & -2.987 & 0.003 & ** \\ \hline
        Month 4 & 2.165 & 90.353 & 0.024 & 0.981 &  \\ \hline
        Month 5 & 61.400 & 85.565 & 0.718 & 0.473 &  \\ \hline
        Month 6 & 252.948 & 86.461 & 2.926 & 0.003 & ** \\ \hline
        Month 7 & 390.221 & 89.391 & 4.365 & 0.000 & *** \\ \hline
        Month 8 & 352.343 & 89.778 & 3.925 & 0.000 & *** \\ \hline
        Month 9 & 312.878 & 72.859 & 4.294 & 0.000 & *** \\ \hline
        Month 10 & 268.331 & 70.165 & 3.824 & 0.000 & *** \\ \hline
        Month 11 & 89.584 & 76.008 & 1.179 & 0.239 &  \\ \hline
        Month 12 & 133.716 & 102.693 & 1.302 & 0.193 &  \\ \hline
        Openness Index & 41.082 & 13.676 & 3.004 & 0.003 & ** \\ \hline
    \end{tabular}
    \caption{Larceny Openness Index Model Results}
    \label{table:larceny-open}
\end{table}

\clearpage

\section*{Appendix C}

\begin{figure}[h]
    \centering
    \includegraphics[scale = 0.5]{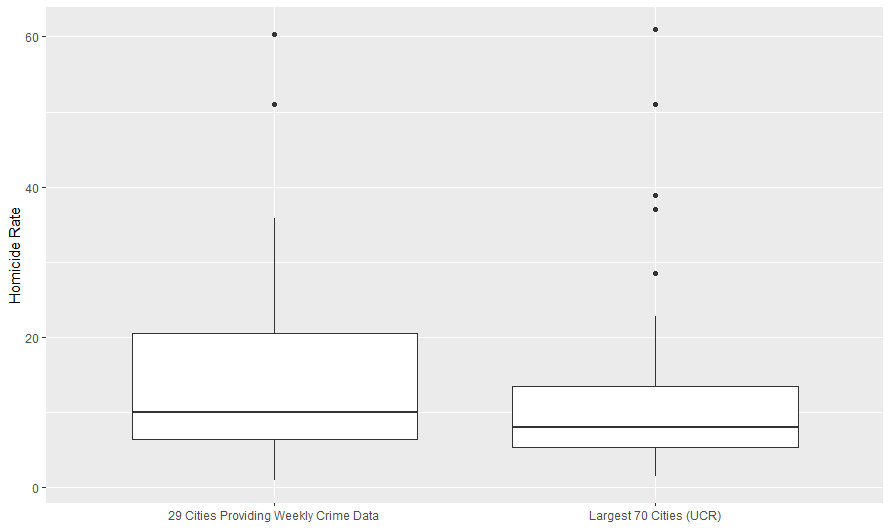}
    \caption{Distribution of City Homicide Rates from Our 2018 Data Compared to 2018 UCR Data}
    \label{fig:hom-ev}
\end{figure}

\begin{figure}[h]
    \centering
    \includegraphics[scale = 0.5]{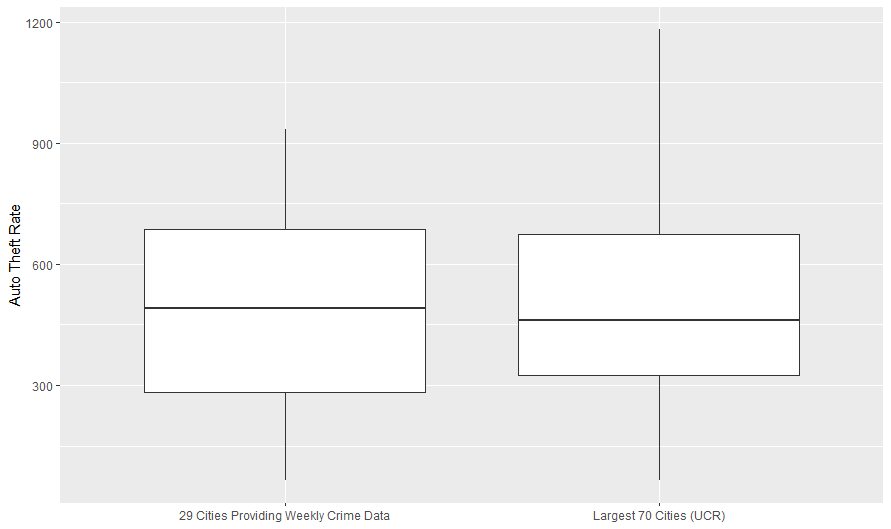}
    \caption{Distribution of City Auto Theft Rates from Our 2018 Data Compared to 2018 UCR Data}
    \label{fig:auto-ev}
\end{figure}

\begin{figure}[h]
    \centering
    \includegraphics[scale = 0.5]{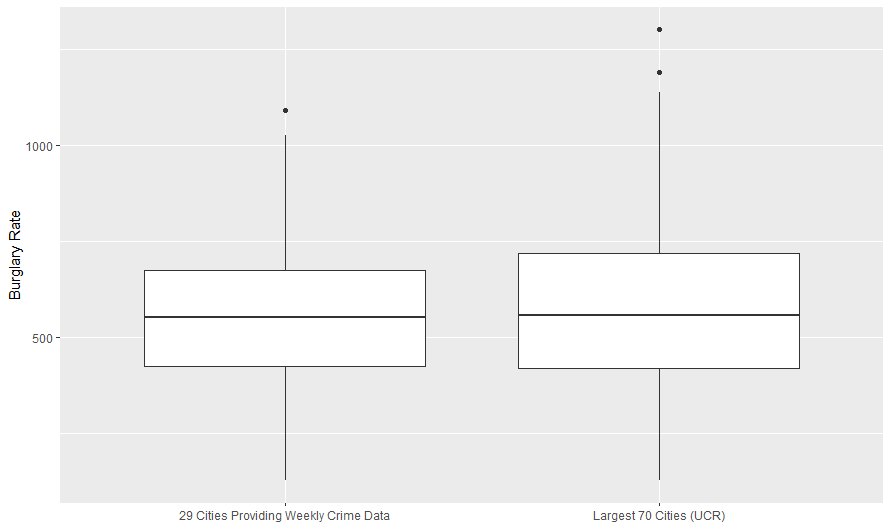}
    \caption{Distribution of City Burglary Rates from Our 2018 Data Compared to 2018 UCR Data}
    \label{fig:burg-ev}
\end{figure}

\begin{figure}[h]
    \centering
    \includegraphics[scale = 0.5]{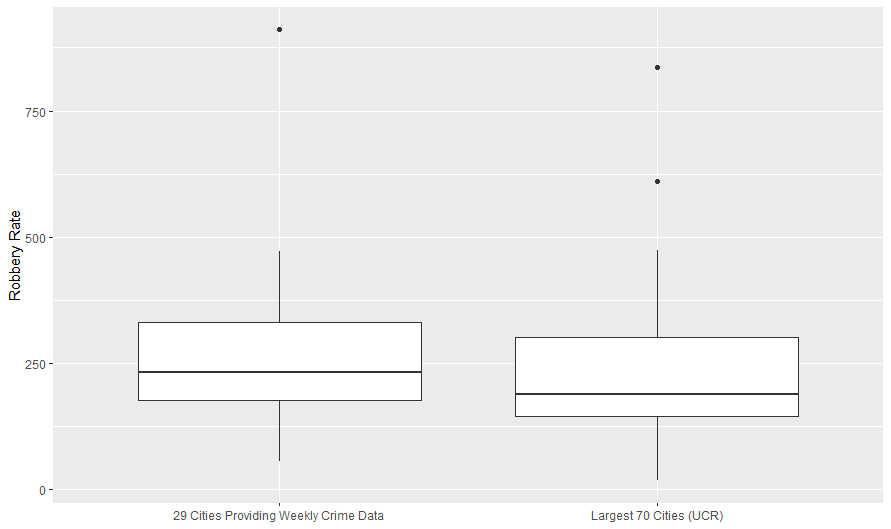}
    \caption{Distribution of City Robbery Rates from Our 2018 Data Compared to 2018 UCR Data}
    \label{fig:rob-ev}
\end{figure}

\begin{figure}
    \centering
    \includegraphics[scale = 0.5]{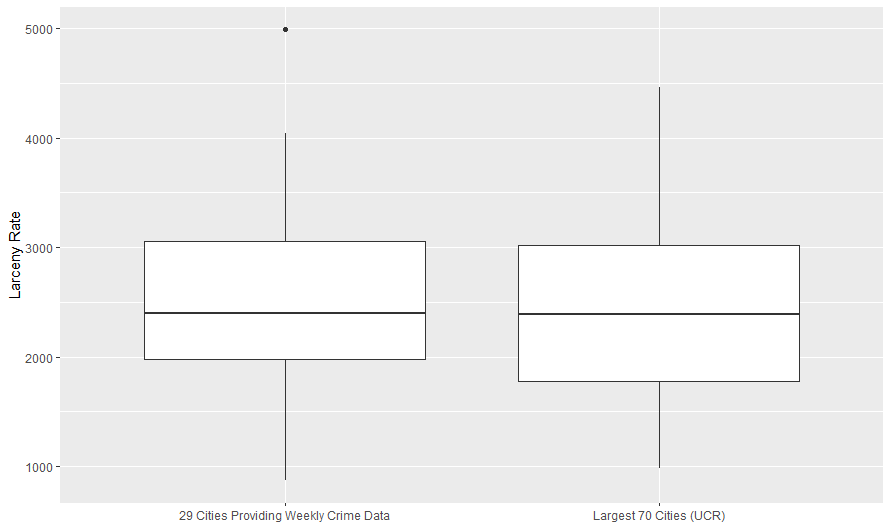}
    \caption{Distribution of City Larceny Rates from Our 2018 Data Compared to 2018 UCR Data}
    \label{fig:larceny-ev}
\end{figure}

\end{document}